\pdfoutput=1

\documentclass[reprint,amsmath,amssymb,floatfix,aps,longbibliography]{revtex4-1}
\usepackage{graphics,graphicx,float}

\usepackage[pdftex,
            pdfauthor={Steven A. Frank},
            pdftitle={},
            pdfsubject={Natural selection. IV. The Price equation}]{hyperref} 

\def\citeyear{\citep}
\def\autocite{\citep}

\newcommand{\zbar}{\bar{z}}
\newcommand{\sz}{z^*}

\newcommand{\wbar}{\bar{w}}

\newcommand{\cov}{{\hbox{\rm Cov}}}
\newcommand{\var}{{\hbox{\rm Var}}}
\newcommand{\bz}{\mathbf{z}}
\newcommand{\bzz}{{-\mskip-12.5mu\mathbf{z}}}

\newcommand{\dq}{\mathbf{\GD q}}
\newcommand{\dqh}{\mathbf{\GD\hat{q}}}
\newcommand{\norm}[1]{\lVert#1\rVert}
\newcommand{\dns}{\GD_\text{\tiny S}\mskip.5mu}
\newcommand{\dne}{\GD_\text{\tiny E}\mskip.5mu}

\newcommand{\Ga}{\alpha}
\newcommand{\Gb}{\beta}

\newcommand{\GD}{\Delta}
\newcommand{\Ge}{\epsilon}

\newcommand{\Gth}{\theta}
\newcommand{\Gf}{\phi}

\DeclareMathOperator{\E}{E}

\newcommand{\Eq}[1]{Eq.~(\ref{eq:#1})}

\newcommand{\Fig}[1]{Fig.~\ref{fig:#1}}

\newcommand{\boldrule}{\hrule height 1.2pt}
\newcommand{\noterule}{\medskip\boldrule\medskip}	% for notes

\newcount\BoxNum \BoxNum 1\relax
\makeatletter
\newcommand{\boxlabel}[1]{%
  \protected@write \@auxout {}{\string \newlabel {box:#1}{{\the\BoxNum}{\thepage}{\noexpand\relax}%
  	{\@ifundefined{hyper@@anchor}{\relax}{box.\the\BoxNum}}%
  	{}}}%
  \@ifundefined{hyper@@anchor}{}{\hypertarget{box.\the\BoxNum}{}}%
  \advance\BoxNum 1\relax}
\makeatother
\newcommand{\Boxx}[1]{Box~\ref{box:#1}}
\newcommand{\BoxLabel}{Box~\the\BoxNum}

\begin{document}

\title{Natural selection. IV. The Price equation}

\author{Steven A.\ Frank}
\email[email: ]{safrank@uci.edu}
\homepage[homepage: ]{http://stevefrank.org}
\affiliation{Department of Ecology and Evolutionary Biology, University of California, Irvine, CA 92697--2525  USA}

\begin{abstract}

The Price equation partitions total evolutionary change into two components.  The first component provides an abstract expression of natural selection. The second component subsumes all other evolutionary processes, including changes during transmission.  The natural selection component is often used in applications.  Those applications attract widespread interest for their simplicity of expression and ease of interpretation.  Those same applications attract widespread criticism by dropping the second component of evolutionary change and by leaving unspecified the detailed assumptions needed for a complete study of dynamics.  Controversies over approximation and dynamics have nothing to do with the Price equation itself, which is simply a mathematical equivalence relation for total evolutionary change expressed in an alternative form.  Disagreements about approach have to do with the tension between the relative valuation of abstract versus concrete analyses.  The Price equation's greatest value has been on the abstract side, particularly the invariance relations that illuminate the understanding of natural selection.  Those abstract insights lay the foundation for applications in terms of kin selection, information theory interpretations of natural selection, and partitions of causes by path analysis. I discuss recent critiques of the Price equation by Nowak and van Veelen\footnote{\href{http://dx.doi.org/10.1111/j.1420-9101.2012.02498.x}{doi:\ 10.1111/j.1420-9101.2012.02498.x} in \textit{J. Evol. Biol.}}\footnote{Part of the Topics in Natural Selection series. See \Boxx{preface}.}.

\end{abstract}

\maketitle

\begin{quote}
The heart and soul of much mathematics consists of the fact that the ``same'' object can be presented to us in different ways. Even if we are faced with the simple-seeming task of ``giving'' a large number, there is no way of doing this without also, at the same time, ``giving'' a hefty amount of extra structure that comes as a result of the way we pin down---or the way we present---our large number. If we write our number as 1729 we are, sotto voce, ordering a preferred way of ``computing it'' (add one thousand to seven hundreds to two tens to nine). If we present it as $1 + 12^3$ we are recommending another mode of computation, and if we pin it down---as Ramanujuan did---as the first number expressible as a sum of two cubes in two different ways, we are being less specific about how to compute our number, but have underscored a characterizing property of it within a subtle diophantine arena.~$\ldots$

This issue has been with us, of course, forever: the general question of \emph{abstraction,} as separating what we want from what we are presented with. It is neatly packaged in the Greek verb \emph{aphairein,} as interpreted by Aristotle in the later books of the \emph{Metaphysics} to mean simply \emph{separation}: if it is \emph{whiteness} we want to think about, we must somehow separate it from \emph{white horse,} \emph{white house,} \emph{white hose,} and all the other white things that it invariably \emph{must} come along with, in order for us to experience it at all \autocite[pp.~222--223]{mazur08when}.\vskip10pt

Somewhere $\ldots$ between the specific that has no meaning and the general that has no content there must be, for each purpose and at each level of abstraction, an optimum degree of generality \autocite[pp.~197--198]{boulding56general}.

\end{quote}

\section*{Introduction}

Evolutionary theory analyzes the change in phenotype over time. We may interpret \emph{phenotype} broadly to include organismal characters, variances of characters, correlations between characters, gene frequency, DNA sequence---essentially anything we can measure. 

How does a phenotype influence its own change in frequency or the change in the frequencies of correlated phenotypes? Can we separate that phenotypic influence from other evolutionary forces that also cause change?  The association of a phenotype with change in frequency, separated from other forces that change phenotype, is one abstract way to describe \emph{natural selection.} The Price equation is that kind of abstract separation.  

Do we really need such abstraction, which may seem rather distant and vague?  Instead of wasting time on such things as the abstract essence of natural selection, why not get down to business and analyze real problems?  For example, we may wish to know how the evolutionary forces of mutation and selection interact to determine biological pattern.  We could make a model with genes that have phenotypic effects, selection that acts on those phenotypes to change gene frequency, and mutation that changes one gene into another.  We could do some calculations, make some predictions about, for example, the frequency of deleterious mutations that cause disease,

\begin{figure}[H]
\begin{minipage}{\hsize}
\parindent=15pt
\noterule
{\bf \noindent\BoxLabel. Topics in the theory of natural selection}
\noterule
\noindent This article is part of a series on natural selection.  Although the theory of natural selection is simple, it remains endlessly contentious and difficult to apply.  My goal is to make more accessible the concepts that are so important, yet either mostly unknown or widely misunderstood.  I write in a nontechnical style, showing the key equations and results rather than providing full derivations or discussions of mathematical problems.  Boxes list technical issues and brief summaries of the literature.  
\noterule
\end{minipage}
\end{figure}
\boxlabel{preface}

\noindent and compare those predictions to observations.  All clear and concrete, without need of any discussion of the essence of things.

However, we may ask the following. Is there some reorientation for the expression of natural selection that may provide subtle perspective, from which we can understand our subject more deeply and analyze our problems with greater ease and greater insight?  My answer is, as I have mentioned, that the Price equation provides that sort of reorientation. To argue the point, I will have to keep at the distinction between the concrete and the abstract, and the relative roles of those two endpoints in mature theoretical understanding.  

Several decades have passed since Price's \citeyear{price70selection,price72extension} original articles. During that span, published claims, counter-claims and misunderstandings have accumulated to the point that it seems worthwhile to revisit the subject.  On the one hand, the Price equation has been applied to numerous practical problems, and has also been elevated by some to almost mythical status, as if it were the ultimate path to enlightenment for those devoted to evolutionary study (\Boxx{literature}). 

On the other hand, the opposition has been gaining adherents who boast the sort of disparaging anecdotes and slogans that accompany battle. In a recent book, \textcite{nowak11supercooperators:} counter
\begin{quote}
The Price equation did not, however, prove as useful as [Price and Hamilton] had hoped. It turned out to be the mathematical equivalent of a tautology.~$\ldots$ If the Price equation is used instead of an actual model, then the arguments hang in the air like a tantalizing mirage. The meaning will always lie just out of the reach of the inquisitive biologist. This mirage can be seductive and misleading. The Price equation can fool people into believing that they have built a mathematical model of whatever system they are studying. But this is often not the case. Although answers do indeed seem to pop out of the equation, like rabbits from a magician's hat, nothing is achieved in reality.
\end{quote}

\textcite{nowak11supercooperators:} approvingly quote \textcite{vanveelen12group} with regard to calling the Price equation a \emph{mathematical tautology}. \textcite{vanveelen12group} emphasize the point by saying that the Price equation is like soccer/football star Johan Cruyff's quip about the secret of success: ``You always have to make sure that you score one goal more than your opponent.'' The statement is always true, but provides no insight.  \textcite{nowak11supercooperators:} and \textcite{vanveelen12group} believe their arguments demonstrate that the Price equation is true in the same trivial sense, and they call that trivial type of truth a \emph{mathematical tautology.}  Interestingly, magazines, online articles, and the scientific literature have for several years been using the phrase \emph{mathematical tautology} for the Price equation, although \textcite{nowak11supercooperators:} and \textcite{vanveelen12group} do not provide citations to previous literature.  

As far as I know, the first description of the Price equation as a mathematical tautology was in \textcite{frank95george}.  I used the phrase in the sense of the epigraph from Mazur, a formal equivalence between different expressions of the same object. Mathematics and much of statistics are about formal equivalences between different expressions of the same object. For example, the Laplace transform changes a mathematical expression into an alternative form with the same information, and analysis of variance decomposes the total variance into a sum of component variances. For any mathematical or statistical equivalence, value depends on enhanced analytical power that eases further derivations and calculations, and on the ways in which previously hidden relations are revealed. 

In light of the contradictory points of view, the main goal of this article is to sort out exactly what the Price equation is, how we should think about it, and its value and limitations in reasoning about evolution. Subsequent articles will show the Price equation in action, applied to kin selection, causal analysis in evolutionary models, and an information perspective of natural selection and Fisher's fundamental theorem.

\section*{Overview}

The first section derives the Price equation in its full and most abstract form.  That derivation allows us to evaluate the logical status of the equation in relation to various claims of fundamental flaw.  The equation survives scrutiny. It is a mathematical relation that expresses the total amount of evolutionary change in an alternative and mathematically equivalent way.  That equivalence provides insight into aspects of natural selection and also provides a guide that, in particular applications, often leads to good approaches for analysis. 

The second section contrasts two perspectives of evolutionary analysis.  In standard models of evolutionary change, one begins with the initial population state and the rules of change. The rules of change include the fitness of each phenotype and the change in phenotype between ancestor and descendant.  Given the initial state and rules of change, one deduces the state of the changed population.  Alternatively, one may have data on the initial population state, the changed population state, and the ancestor-descendant relations that map entities from one population to the other.  Those data may be reduced to the evolutionary distance between two populations, providing inductive information about the underlying rules of change.  Natural populations have no intrinsic notion of fitness or rules of change. Instead, they inductively accumulate information. The Price equation includes both the standard deductive model of evolutionary change and the inductive model by which information accumulates in relation to the evolutionary distance between populations.  

The third and fourth sections discuss the Price equation's abstract properties of invariance and recursion. The invariance properties include the information theory interpretation of natural selection. Recursion provides the basis for analyzing group selection and other models of multilevel selection.

The fifth section relates the Price equation to various expressions that have been used throughout the history of evolutionary theory to analyze natural selection.  The most common form describes natural selection by the covariance between phenotype and fitness or by the covariance between genetic breeding value and fitness. The covariance expression is one part of the Price equation that, when used alone, describes the natural selection component of total evolutionary change. The essence of those covariance forms arose in the early studies of population and quantitative genetics, have been used extensively during much of the modern history of animal breeding, and began to receive more mathematical development in the 1960s and 1970s.  Recent critiques of the Price equation focus on the same covariance expression that has been widely used throughout the history of population and quantitative genetics to analyze natural selection and to approximate total evolutionary change.

The sixth section returns to the full abstract form of the equation.  I compare a few variant expressions that have been promoted as improvements on the original Price equation. Variant forms are indeed helpful with regard to particular abstract problems or particular applications.  However, most variants are simply minor rearrangements of the mathematical equivalence for total evolutionary change given by the original Price equation.  The recent extension by \textcite{kerr09generalization} does provide a slightly more general formulation by expanding the fundamental set mapping that defines Price's approach.  The set mapping basis for the Price equation deserves more careful study and further mathematical work.

The seventh section analyzes various flaws that have been ascribed to the Price equation. For example, the Price equation in its most abstract form does not contain enough information to follow evolutionary dynamics through multiple rounds of natural selection.  By contrast, classical dynamic models of population genetics are sufficient to follow change through time.  Much has been  

\begin{figure}[H]
\begin{minipage}{\hsize}
\parindent=15pt
\noterule
{\bf \noindent\BoxLabel. Price equation literature}
\noterule
\noindent A large literature introduces and reviews the Price equation.  I list some key references that can be used to get started \autocite{hamilton75innate,frank95george,frank97the-price,grafen02a-first,page02unifying,andersen04population,rice04evolutionary,okasha06evolution,gardner08the-price}.

Diverse applications have been developed with the Price equation.  I list a few examples \autocite{hamilton70selfish,wade85soft,frank90the-distribution,queller92a-general,queller92quantitative,michod97evolution,michod97cooperation,frank98foundations,fox06using,day06insights,grafen07the-formal,alizon09the-price}.  

Quantitative genetics theory often derives from the covariance expression given by \textcite{robertson66a-mathematical}, which is a form of the covariance term of the Price equation.  The basic theory can be found in textbooks \autocite{falconer96introduction,charlesworth10elements}. Much of the modern work can be traced through the widely cited article by \textcite{lande83the-measurement}. 

\textcite{harman10the-price} provides an interesting overview of Price's life and evokes an Olympian sense of the power and magic of the Price equation.  See \textcite{schwartz00death} for an alternative biographical sketch.
\noterule
\end{minipage}
\end{figure}
\boxlabel{literature}

\noindent made of this distinction with regard to dynamic sufficiency.  The distinction arises from the fact that classical dynamics in population genetics makes more initial assumptions than the abstract Price equation.  It must be true that all mathematical equivalences for total evolutionary change have the same dynamic status given the same initial assumptions. Each additional well-chosen assumption typically enhances the specificity and reduces the scope and generality of the analysis.  The epigraph from Boulding emphasizes that the degree of specificity versus generality is an explicit choice of the analyst with respect to initial assumptions. 

The Discussion considers the value and limitations of the Price equation in relation to recent criticisms by Nowak and van Veelen.  The critics confuse the distinct roles of general abstract theory and concrete dynamical models for particular cases.   The enduring power of the Price equation arises from the discovery of essential invariances in natural selection. For example, kin selection theory expresses biological problems in terms of relatedness coefficients. Relatedness measures the association between social partners. The proper measure of relatedness identifies distinct biological scenarios with the same (invariant) evolutionary outcome. Invariance relations provide the deepest insights of scientific thought.

\section*{The Price equation}

\begin{quote}
The mathematics given here applies not only to genetical selection but to selection in general. It is intended mainly for use in deriving general relations and constructing theories, and to clarify understanding of selection phenomena, rather than for numerical calculation \autocite[p.~485]{price72extension}.
\end{quote}

I have emphasized that the Price equation is a mathematical equivalence.  The equation focuses on separation of total evolutionary change into a part attributed to selection and a remainder term.  That separation provides an abstraction of the nature of selection. As Price wrote sometime around 1970 but published posthumously in \textcite{price95the-nature}: ``Despite the pervading importance of selection in science and life, there has been no abstraction and generalization from genetical selection to obtain a general selection theory and general selection mathematics.''  

It is useful first to consider the Price equation in this most abstract form. I follow my earlier derivations \autocite{frank95george,frank97the-price,frank98foundations,frank09natural}, which differ little from the derivation given by \textcite{price72extension} when interpreted in light of \textcite{price95the-nature}.

The abstract expression can best be thought of in terms of mapping items between two sets \autocite{frank95george,price95the-nature}. In biology, we usually think of an ancestral population at some time and a descendant population at a later time.  Although there is no need to have an ancestor-descendant relation, I will for convenience refer to the two sets as ancestor and descendant.  What does matter is the relations between the two sets, as follows.

\subsection*{Definitions}

The full abstract power of the Price equation requires adhering strictly to particular definitions.  The definitions arise from the general expression of the relations between two sets.

Let $q_i$ be the frequency of the $i$th type in the ancestral population.  The index $i$ may be used as a label for any sort of property of things in the set, such as allele, genotype, phenotype, group of individuals, and so on.  Let $q'_i$ be the frequencies in the descendant population, defined as the fraction of the descendant population that is derived from members of the ancestral population that have the label $i$.  Thus, if $i=2$ specifies a particular phenotype, then $q'_2$ is not the frequency of the phenotype $i=2$ among the descendants.  Rather, it is the fraction of the descendants derived from entities with the phenotype $i=2$ in the ancestors.  One can have partial assignments, such that a descendant entity derives from more than one ancestor, in which case each ancestor gets a fractional assignment of the descendant.  The key is that the $i$ indexing is always with respect to the properties of the ancestors, and descendant frequencies have to do with the fraction of descendants derived from particular ancestors.  

Given this particular mapping between sets, we can specify a particular definition for fitness.  Let $q'_i = q_i(w_i/\wbar)$, where $w_i$ is the fitness of the $i$th type and $\wbar = \sum q_iw_i$ is average fitness.  Here, $w_i/\wbar$ is proportional to the fraction of the descendant population that derives from type $i$ entities in the ancestors.  

Usually, we are interested in how some measurement changes or evolves between sets or over time.  Let the measurement for each $i$ be $z_i$.  The value $z$ may be the frequency of a gene, the squared deviation of some phenotypic value in relation to the mean, the value obtained by multiplying measurements of two different phenotypes of the same entity, and so on.  In other words, $z_i$ can be a measurement of any property of an entity with label, $i$.  The average property value is $\zbar=\sum q_iz_i$, where this is a population average.  

The value $z'_i$ has a peculiar definition that parallels the definition for $q'_i$.  In particular, $z'_i$ is the average measurement of the property associated with $z$ among the descendants derived from ancestors with index $i$.  The population average among descendants is $\zbar'=\sum q'_iz'_i$.

The Price equation expresses the total change in the average property value, $\GD\zbar=\zbar'-\zbar$, in terms of these special definitions of set relations.  This way of expressing total evolutionary change and the part of total change that can be separated out as selection is very different from the usual ways of thinking about populations and evolutionary change.  The derivation itself is very easy, but grasping the meaning and becoming adept at using the equation is not so easy.  

I will present the derivation in two stages.  The first stage makes the separation into a part ascribed to selection and a part ascribed to property change that covers everything beyond selection.  The second stage retains this separation, changing the notation into standard statistical expressions that provide the form of the Price equation commonly found in the literature.  I follow with some examples to illustrate how particular set relations are separated into selection and property change components. The next section considers two distinct interpretations of the Price equation in relation to dynamics.

\subsection*{Derivation: separation into selection and property value change}

We use $\GD q_i = q'_i-q_i$ for frequency change associated with selection, and $\GD z_i = z'_i-z_i$ for property value change. Both expressions for change depend on the special set relation definitions given above. 

We are after an alternative expression for total change, $\GD\zbar$. Thus,
\begin{align*}
  \GD\zbar&= \zbar'-\zbar\\
          &= \sum q'_iz'_i - \sum q_iz_i\\
          &= \sum q'_i(z'_i-z_i) +\sum q'_iz_i - \sum q_iz_i\\
          &= \sum q'_i(\GD z_i) + \sum(\GD q_i)z_i.
\end{align*}
Switching the order of the terms on the right side of the last line yields
\begin{equation}\label{eq:PriceSum}
  \GD\zbar = \sum(\GD q_i)z_i + \sum q'_i(\GD z_i),
\end{equation}
a form emphasized by \textcite[eqn~1]{frank97the-price}.  The first term separates the part of total change caused by changes in frequency.  We call this the part caused by selection, because this is the part that arises directly from differential contribution by ancestors to the descendant population \autocite{price95the-nature}.  Because the set mappings define all of the direct attributions of success for each $i$ with respect to the associated properties $z_i$, it is reasonable to separate out this direct component as the abstraction of \emph{selection.} It is of course possible to define other separations.  I discuss one particular alternative later. However, it is hard to think of other separations that would describe selection in a better way at the most abstract and general level of the mappings between two sets.  This first term has also been called the partial evolutionary change caused by natural selection (\Eq{partial}).

The second term describes the part of total change caused by changes in property values.  Recall that $\GD z_i=z'_i-z_i$, and that $z'_i$ is the property value among entities that descend from $i$.  Many different processes may cause descendant property values to differ from ancestral values.  In fact, the assignment of a descendant to an ancestor can be entirely arbitrary, so that there is no reason to assume that descendants should be like ancestors.  Usually, we will work with systems in which descendants do resemble ancestors, but the degree of such associations can be arranged arbitrarily. This term for change in property value encompasses everything beyond selection.  The idea is that selection affects the relative contribution of ancestors and thus the changes in frequencies of representation, but what actually gets represented among the descendants will be subject to a variety of processes that may alter the value expressed by descendants.  

The equation is exact and must apply to every evolutionary system that can be expressed as two sets with certain ancestor-descendant or mapping relations.  It is in that sense that I first used the phrase \emph{mathematical tautology} \autocite{frank95george}.  The nature of separation and abstraction is well described by the epigraph from Mazur at the start of this article.  

\subsection*{Derivation: statistical notation}

\textcite{price72extension} used statistical notation to write \Eq{PriceSum}.  For the first term, by following prior definitions we have
\begin{align*}
  \GD q_i &= q'_i - q_i\\
          &=q_i\frac{w_i}{\wbar} - q_i\\
          &=q_i\left(\frac{w_i}{\wbar}-1\right),
\end{align*}
so that
\begin{equation*}
  \sum (\GD q_i)z_i = \sum q_i\left(\frac{w_i}{\wbar}-1\right)z_i=\cov(w,z)/\wbar,
\end{equation*}
using the standard definition for population covariance.  

For the second term, we have
\begin{equation*}
  \sum q'_i(\GD z_i) = \sum q_i\frac{w_i}{\wbar}(\GD z_i) = \E(w\GD z)/\wbar,
\end{equation*}
where $\E$ means expectation, or average over the full population.  Putting these statistical forms into \Eq{PriceSum} and moving $\wbar$ to the left side for notational convenience yields a commonly published form of the Price equation
\begin{equation}\label{eq:PriceEq}
  \wbar\GD\zbar = \cov(w,z) + \E(w\GD z).
\end{equation}
\textcite{price95the-nature} and \textcite{frank95george} present examples of set mappings expressed in relation to the Price equation.  

\section*{Dynamics: inductive and deductive perspectives}

The Price equation describes evolutionary change between two populations. Three factors express one iteration of dynamical change: initial state, rules of change, and next state.  In the Price equation, the phenotypes, $z_i$, and their frequencies, $q_i$, describe the initial population state. Fitnesses, $w_i$, and property changes, $\GD z_i$, set the rules of change.  Derived phenotypes, $z_i'$, and their frequencies, $q_i'$, express the next population state. 

Models of evolutionary change essentially always analyze forward or deductive dynamics.  In that case, one starts with initial conditions and rules of change and calculates the next state.  Most applications of the Price equation use this traditional deductive analysis.  Such applications lead to predictions of evolutionary outcome given assumptions about evolutionary process, expressed by the fitness parameters and property changes.  

Alternatively, one can take the state of the initial population and the state of the changed population as given. If one also has the mappings between initial and changed populations that connect each entity, $i$, in the initial population to entities in the changed population, then one can calculate (induce) the underlying rules of change.  At first glance, this inductive view of dynamics may seem rather odd and not particularly useful.  Why start with knowledge of the evolutionary sequence of population states and ancestor-descendant relations as given, and inductively calculate fitnesses and property changes?  The inductive view takes the fitnesses, $w_i$, to be derived from the data rather than an intrinsic property of each type.  

The Price equation itself does not distinguish between the deductive and inductive interpretations.  One can specify initial state and rules of change and then deduce outcome. Or one can specify initial state and outcome along with ancestor-descendant mappings, and then induce the underlying rules of change.  It is useful to understand the Price equation in its full mathematical generality, and to understand that any specific interpretation arises from additional assumptions that one brings to a particular problem.  Much of the abstract power of the Price equation comes from understanding that, by itself, the equation is a minimal description of change between populations.

The deductive interpretation of the Price equation is clear.  What value derives from the inductive perspective?  In observational studies of evolutionary change, we only have data on population states.  From those data, we use the inductive perspective to make inferences about the underlying rules of change.  Note that inductive estimates for evolutionary process derive from the amount of change, or distance, between ancestor and descendant populations.  The Price equation includes that inductive, or retrospective view, by expressing the distance between populations in terms of $\GD\zbar$.  I develop that distance interpretation in the following sections.  

Perhaps more importantly, natural selection itself is inherently an inductive process by which information accumulates in populations.  Nature does not intrinsically ``know'' of fitness parameters.  Instead, frequency changes and the mappings between ancestor and descendant are inherent in a population's response to the environment, leading to a sequence of population states, each separated by an evolutionary distance.  That evolutionary distance provides information that populations accumulate inductively about the fitnesses of each phenotype \autocite{frank09natural}.  The Price equation includes both the deductive and inductive perspectives.  We may choose to interpret the equation in either way depending on our goals of analysis.

\section*{Abstract properties: invariance}

The Price equation describes selection by the term $\sum(\GD q_i)z_i= \cov(w,z)/\wbar$. Any instance of evolutionary change that has the same value for this sum has the same amount of total selection.  Put another way, for any particular value for total selection, there is an infinite number of different combinations of frequency changes and character measurements that will add up to the same total value for selection.  All of those different combinations lead to the same value with respect to the amount of selection.  We may say that all of those different combinations are \emph{invariant} with respect to the total quantity of selection.  The deepest insights of science come from understanding what does not matter, so that one can also say exactly what does matter---what is invariant \autocite{feynman67the-character,weyl83symmetry}.

The invariance of selection with respect to transformations of the fitnesses, $w$, and the phenotypes, $z$, that have the same $\cov(w,z)$ means that, to evaluate selection, it is sufficient to analyze this covariance.  At first glance, it may seem contradictory that the covariance, commonly thought of as a linear measure of association, can be a complete description for selection, including nonlinear processes.  Let us step through this issue, first looking at why the covariance is a sufficient expression of selection, and then at the limitations of this covariance expression in evolutionary analysis.

\subsection*{Covariance as a measure of distance: definitions}

Much of the confusion with respect to covariance and variance terms in selection equations arises from thinking only of the traditional statistical usage. In statistics, covariance typically measures the linear association between pairs of observations, and variance is a measure of the squared spread of observations.  Alternatively, covariances and variances provide measures of distance, which ultimately can be understood as measures of information \autocite{frank09natural}.  This section introduces the notation for the geometric interpretation of distance.  The next section gives the main geometric result, and the following section presents some examples.  

The identity $\sum(\GD q_i)z_i= \cov(w,z)/\wbar$ provides the key insight.  It helps to write this identity in an alternative form.  Note from the prior definition $q_i' = q_iw_i/\wbar$ that 
\begin{equation}\label{eq:aveExcess}
  \GD q_i = q'_i - q_i = q_i(w_i/\wbar - 1) = q_ia_i,
\end{equation}
where $a_i=w_i/\wbar - 1$ is Fisher's average excess in fitness, a commonly used expression in population and quantitative genetics \autocite{fisher30the-genetical,fisher41average,crow70an-introduction}.  A value of zero means that an entity has average fitness, and therefore fitness effects and selection do not change the frequency of that entity.  Using the average excess in fitness, we can write the invariant expression for selection as
\begin{equation}\label{eq:freqChange}
  \sum(\GD q_i)z_i= \sum q_ia_iz_i = \cov(w,z)/\wbar.
\end{equation}

We can think of the state of the population as the listing of character states, $z_i$.  Thus we write the population state as $\bz = (z_1,z_2,\ldots)$.  The subscripts run over every different entity in the population, so the vector $\bz$ is a complete description of the entire population.  Similarly, for the frequency fluctuations, $\GD q_i = q_ia_i$, we can write the listing of all fluctuations as a vector, $\dq=(\GD q_1,\GD q_2,\ldots)$.

It is often convenient to use the dot product notation
\begin{equation*}
  \dq \cdot \bz=\sum(\GD q_i)z_i= \cov(w,z)/\wbar
\end{equation*}
in which the dot specifies the sum obtained by multiplying each pair of items from two vectors.  Before turning to some geometric examples in the following section, we need a definition for the length of a vector.  Traditionally, one uses the definition
\begin{equation*}
  \norm{\bz} = \sqrt{\sum z_i^2},
\end{equation*}
in which the length is the square root of the sum of squares, which is the standard measure of length in Euclidean geometry.  

\subsection*{Covariance as a measure of distance: examples}

\begin{figure}[H]
\centering
\includegraphics[width=1.85in]{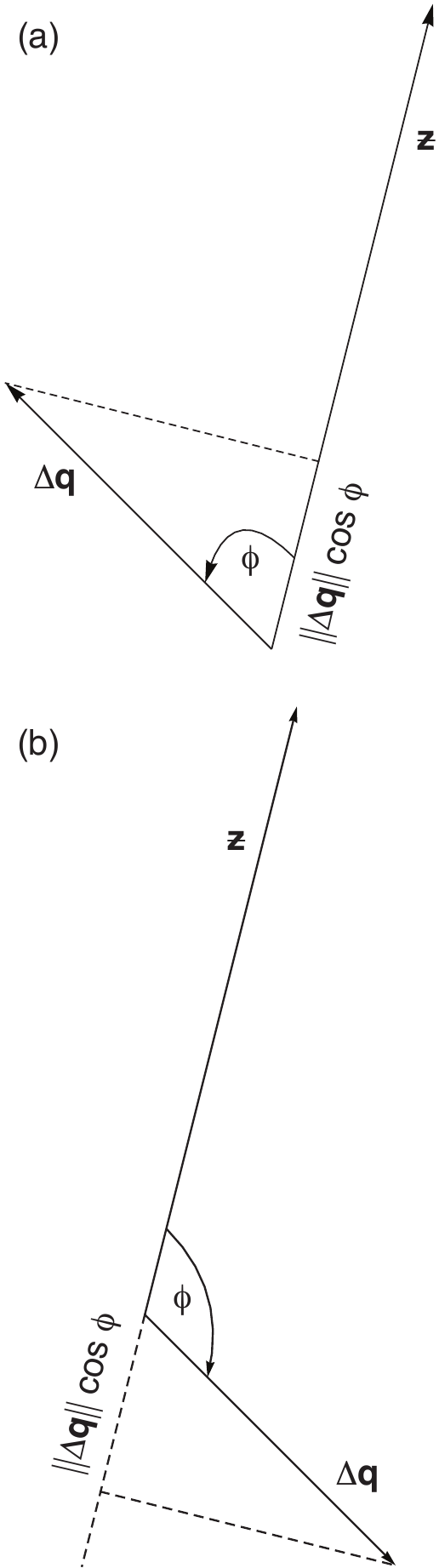}
\caption{Geometric expression of selection.  The plots show the equivalence of the dot product, the geometric expression and the covariance, as given in \Eq{dotGeom}.  For both plots, $\bz=(1,4)$ and $\bzz=\bz/\norm{\bz}=(0.24,0.97)$. The dashed line shows the perpendicular between the pattern of frequency changes derived from fitnesses, $\dq$, and the phenotypic pattern, $\bzz$.  The vertex of the two vectors is at the origin $(0,0)$. The distance from the origin to the intersection of the perpendicular along $\bzz$ is the total amount of selection, $\norm{\dq}\cos\Gf$. (a) The vector of frequency changes that summarize fitness is $\dq=(-0.4,0.4)$. The angle between the vector of frequency changes and the phenotypes is $\Gf=\arccos\left[(\dq \cdot \bzz)/\norm{\dq}\right]$ which, in this example, is 1.03 radians or $59^{\circ}$. In this case, the total selection is $\norm{\dq}\cos\Gf=0.29$. (b) In this plot, $\dq=(0.4,-0.4)$, yielding an angle $\Gf$ of $121^{\circ}$.  The perpendicular intersects the negative projection of the phenotype vector, shown as a dashed line, associated with the negative change by selection of $\norm{\dq}\cos\Gf=-0.29$.
\label{fig:geom2D}
}
\end{figure}

A simple identity relates a dot product to a measure of distance and to covariance selection
\begin{equation}\label{eq:dotGeom}
  \dq \cdot \bz = \norm{\dq}\norm{\bz}\cos\Gf=\cov(w,z)/\wbar,
\end{equation}
where $\Gf$ is the angle between the vectors $\dq$ and $\bz$ (\Fig{geom2D}). If we standardize the character vector $\bzz = \bz/\norm{\bz}$, then the standardized vector has a length of one, $\norm{\bzz}=1$, which simplifies the dot product expression of selection to
\begin{equation*}
  \dq \cdot \bzz = \norm{\dq}\cos\Gf,
\end{equation*}
providing the geometric representation illustrated in \Fig{geom2D}.

The covariance can be expressed as the product of a regression coefficient and a variance term
\begin{equation}\label{eq:varZ}
  \cov(w,z)/\wbar=\Gb_{z w}\var(w)/\wbar=\Gb_{w z}\var(z)/\wbar,
\end{equation}
where the notation $\Gb_{x y}$ describes the regression coefficient of $x$ on $y$ \autocite{price70selection}.  This identity shows that the expression of selection in terms of a regression coefficient and a variance term is equivalent to the geometric expression of selection in terms of distance.  

I emphasize these identities for two reasons.  First, as Mazur stated in the epigraph: ``The heart and soul of much mathematics consists of the fact that the `same' object can be presented to us in different ways.'' If an object is important, such as natural selection surely is, then it pays to study that object from different perspectives to gain deeper insight.  

Second, the appearance of statistical functions, such as the covariance and variance, in selection equations sometimes leads to mistaken conclusions.  In the selection equations, it is better to think of the covariance and variance terms arising because they are identities with geometric or other interpretations of selection, rather than thinking of those terms as summary statistics of probability distributions.  The problem with thinking of those terms as statistics of probability distributions is that the variance and covariance are not in general sufficient descriptions for probability distributions.  That lack of sufficiency for probability may lead one to conclude that those terms are not sufficient for a general expression of selection.  However, those covariance and variance terms are sufficient. That sufficiency can be understood by thinking of those terms as identities for distance or measures of information \autocite{frank09natural}.

It is true that in certain particular applications of quantitative genetics or stochastic sampling processes, one does interpret the variances and covariances as summary statistics of probability distributions, usually the normal or Gaussian distribution.  However, it is important to distinguish those special applications from the general selection equations. 

\subsection*{Invariance and information}

For the general selection expression in \Eq{dotGeom}, any transformations that do not affect the net values are invariant with respect to selection.  For example, transformations of the fitnesses and associated frequency changes, $\dq$, are invariant if they leave unchanged the distance expressed by $\dq\cdot\bz=\cov(w,z)/\wbar$.  Similarly, changes in the pattern of phenotypes are invariant to the 

\noindent extent that they leave $\dq\cdot\bz$ unchanged. These invariance properties of selection, measured as distance, may not appear very interesting at first glance.  They seem to be saying that the outcome is the outcome.  However, the history of science suggests that studying the invariant properties of key expressions can lead to insight.  

Few authors have developed an interest in the invariant qualities of selection.  \textcite{fisher30the-genetical} initiated discussion with his fundamental theorem of natural selection, a special case of \Eq{dotGeom} \autocite{frank97the-price}. Although many authors commented on the fundamental theorem, most articles did not analyze the theorem with respect to its essential mathematical insights about selection. \textcite{ewens92an-optimizing} reviewed the few attempts to understand the mathematical basis of the theorem and its invariant quantities. \textcite{frank09natural} tied the theorem to Fisher information \autocite{frieden01population,frieden04science}, hinting at an information theory interpretation that arises from the fundamental selection equation of \Eq{dotGeom}.  

In spite of the importance of selection in many fields of science, the potential interpretation of \Eq{dotGeom} with respect to invariants of information theory has hardly been developed.  I briefly outline the potential connections here \autocite{frank09natural}. I develop this information perspective of selection in a later article, along with Fisher's fundamental theorem.

To start, define the partial change in phenotype caused by natural selection as
\begin{equation}\label{eq:partial}
  \dns\zbar = \dq\cdot\bz = \cov(w,z)/\wbar.
\end{equation}
The concept of a partial change caused by natural selection arises from Fisher's fundamental theorem \autocite{fisher30the-genetical,price72fishers,ewens89an-interpretation,frank92fishers}. With this definition, we can use eqns \ref{eq:dotGeom} and \ref{eq:varZ} to write
\begin{equation}\label{eq:ds1}
  \dns\zbar = \Gb_{z w}\var(w)/\wbar =  \wbar\Gb_{z w}\var(w/\wbar).
\end{equation}
From \Eq{aveExcess}, we have the definition for the average excess in fitness $a_i=w_i/\wbar-1$.  Thus, we can expand the expression for the variance in fitness as
\begin{equation*}
  \var(w/\wbar) = \sum q_i\left(\frac{w_i}{\wbar}-1\right)^2 =\sum q_ia_i^2.
\end{equation*}
From \Eq{aveExcess}, we also have the change in frequency in terms of the average excess, $\GD q_i = q_ia_i$, and equivalently, $a_i = \GD q_i/q_i$, thus
\begin{align*}
  \var(w/\wbar) &= \sum q_i\left(\frac{\GD q_i}{q_i}\right)^2\\
             &= \sum \left(\frac{\GD q_i}{\sqrt{q_i}}\right)^2\\
             &= \dqh\cdot\dqh,
\end{align*}
where $\GD\hat{q}_i=\GD q_i/\sqrt{q_i}$ is a standardized fluctuation in frequency, and $\dqh$ is the vector of standardized fluctuations.  These alternative forms simply express the variance in fitness in different ways.  The interesting result follows from the fact that
\begin{equation*}
  \var(w/\wbar)=\dqh\cdot\dqh = F(\dqh)
\end{equation*}
is the Fisher information,  $F$, in the frequency fluctuations, $\dqh$.  Fisher information is a fundamental quantity in information theory, Bayesian analysis, likelihood theory and the informational foundations of statistical inference.  Fisher information is a variant form of the more familiar Shannon and Kullback-Leibler information measures, in which the Fisherian form expresses changes in information.  

Once again, we have a simple identity.  Although it is true that Fisher information is just an algebraic rearrangement of the variance in fitness, some insight may be gained by relating selection to information.  The variance form calls to mind a statistical description of selection or a partial description of a probability distribution. The Fisher information form suggests a relation between natural selection and the way in which populations accumulate information \autocite{frank09natural}.

We may now write our fundamental expression for selection as
\begin{equation*}
  \dns\zbar = \wbar\Gb_{zw}\,F(\dqh).
\end{equation*} 
We may read this expression for selection as: the change in mean character value caused by natural selection, $\dns\zbar$, is equal to the total Fisher information in the frequency fluctuations, $F$, multiplied the scaling $\Gb$ that describes the amount of the potential information that the population captures when expressed in units of phenotypic change. In other words, the distance $\dns\zbar$ measures the informational gain by the population caused by natural selection.   

The invariances set by this expression may be viewed in different ways. For example, the distance of evolutionary change by selection, $\dns\zbar$, is invariant with respect to many different combinations of frequency fluctuations, $\dqh$, and scalings between phenotype and fitness.  Similarly, any transformations of frequency fluctuations that leave the measure of information, $F(\dqh)$, invariant do not alter the scaled change in phenotype caused by natural selection. The full implications remain to be explored.  

{\bf NOTE ADDED AFTER PUBLICATION OF THE JOURNAL ARTICLE.}  I have oversimplified a bit in this section, with the aim to keep the presentation brief.  The proper expression of Fisher information is $F(\dqh)(\GD\Gth)^2$, where $\GD\Gth$ is the scale of change over which population differences are measured, typically taken as $\GD\Gth\rightarrow0$ \autocite{frank09natural}.  

\subsection*{Summary of selection identities}

The various identities for the part of total evolutionary change caused by selection include
\begin{align}
  \dns\zbar &= \cov(w,z)/\wbar \notag\\
                   &= \wbar\Gb_{zw}\var(w/\wbar) \notag\\
                   &= \dq\cdot\bz \notag\\
                   &= \norm{\dq}\norm{\bz}\cos\Gf \notag\\
                   &= \wbar\Gb_{zw}(\dqh\cdot\dqh) \notag\\
                   &= \wbar\Gb_{zw}\,F(\dqh). \label{eq:altCov}
\end{align}
These forms show the equivalence of the statistical, geometrical and informational expressions for natural selection.  These general abstract forms make no assumptions about the nature of phenotypes and the patterns of frequency fluctuations caused by differential fitness.  The phenotypes may be squared deviations so that the average is actually a variance, or the product of measurements on different characters leading to measures of association, or any other nonlinear combination of measurements.  Thus, there is nothing inherently linear or restrictive about these expressions.  

\subsection*{Selection versus evolution}

The previous sections discussed the part of evolutionary change caused by selection.  The full Price equation (\Eq{PriceEq}) gives a complete and exact expression of total change, repeated here as
\begin{equation}\label{eq:PE2}
  \GD\zbar = \cov(w,z)/\wbar + \E(w\GD z)/\wbar
\end{equation}
or in terms of the dot product notation as
\begin{equation}\label{eq:PEbold}
  \GD\zbar = \dq\cdot\bz \mskip3mu + \mskip3mu\mathbf{q'}\cdot\mathbf{\GD z}.
\end{equation}
The full change in the phenotype is the sum of the two terms, which we may express in symbols as
\begin{equation*}
  \GD\zbar = \dns\zbar + \dne\zbar.
\end{equation*}
\textcite{fisher30the-genetical} called the term $\dne\zbar$ the change caused by the environment \autocite{frank92fishers}.  However, the word \emph{environment} often leads to confusion. The proper interpretation is that $\dne\zbar$ encompasses \emph{everything} not included in the expression for selection.  The term is \emph{environmental} only in the sense that it includes all those forces \emph{external} to the particular definition of the selective forces for a particular problem.  

The $\dne$ term is sometimes associated with changes in transmission \autocite{frank95george,frank97the-price,frank12natural,okasha06evolution}.  This interpretation arises because $\E(w\GD z)$ is the fitness weighted changes in character value between ancestor and descendant.  One may think of changes in character values as changes during transmission.  

It is important to realize that \emph{everything} truly means every possible force that might arise and that is not accounted for by the particular expression for selection.  Lightning may strike. New food sources may appear.  The Price equation in its general and abstract form is a mathematical identity---what I previously called a \emph{mathematical tautology} \autocite{frank95george}. 

In applications, one considers how to express $\dne\zbar$, or one searches for ways to formulate the problem so that $\dne\zbar$ is zero or approximately zero.  This article is not about particular applications. Here, I simply note that when one works with Fisher's breeding value as $z$, then near equilibria (fixed points), one typically obtains $\GD z\rightarrow 0$ and thus $\E(w\GD z)\rightarrow0$.  In other cases, the search for a good way to express a problem means finding a form of character measurement that defines $z$ such that characters tend to remain stable over time, so that $\GD z\rightarrow 0$ and thus $\E(w\GD z)\rightarrow0$.  For applications that emphasize calculation of complex dynamics rather than a more abstract conceptual analysis of a problem, methods other than the Price equation often work better.

\section*{Abstract properties: recursion and group selection}

\begin{quote}
To iterate is human, to recurse, divine \autocite{coplien98to-iterate}.
\end{quote}

\noindent Essentially all modern discussions of multilevel selection and group selection derive from \textcite{price72extension}, as developed by \textcite{hamilton75innate}.  Price and Hamilton noted that the Price equation can be expanded recursively to represent nested levels of analysis, for example, individuals living in groups.  

Start with the basic Price equation as given in \Eq{PE2}.  The left side is the total change in average phenotype, $\zbar$.  The second term on the right side includes the terms $\GD z_i$ in $\E(w\GD z) = \sum{q_i w_i\GD z_i}$. 

Recall that in defining $z_i$, we specified the meaning of the index $i$ to be any sort of labeling of set members, subject to minimal consistency requirements.  We may, for example, label all members of a group by $i$, and measure $z_i$ as some property of the group.  If the index $i$ itself represents a set, then we may consider the members of that set.  For example, $z_{ij}$ may be the $j$th member of the $i$th set, or we may say, the $i$th group.  In the abstract mathematical expression, there is no need to think of the $i$th group as having any spatial or biological meaning.  However, we may consider $i$ as a label for spatially defined groups if we wish to do so.  

With $i$ defining a group, we may analyze the selection and evolution of that $i$th group.  The term $\GD z_i$ becomes the average change in the $z$ measure for the $i$th group, composed of members with values $z_{ij}$.  The  terms $z'_{ij}$ are the average property values of the descendants of the $j$th entity in the $i$th group. The descendant entities that derive from the $i$th group do not have to form any sort of group or other meaningful structuring, just as the original $i$ labeling does not have to refer to group structuring in the ancestors.  However, we may if we wish consider descendants of $i$ as retaining some sense of the ancestral grouping. 

Because $z_i$ represents an averaging over the entities $j$ in the $i$th group, we are assuming the notational equivalence $\GD z_i = \GD\zbar_i$.  From that point of view, for each group $i$ we may from \Eq{PE2} express the change in the group mean by thinking of each group as a separate set or population, yielding for each $i$ the expression
\begin{equation*}
  \GD z_i = \GD\zbar_i = \cov(w_i,z_i)/\wbar_i + \E(w_i\GD z_i)/\wbar_i.
\end{equation*}
We may substitute this expression for each $i$ into the $\E(w\GD z) =  \sum{q_i w_i\GD z_i}$ term on the right side of \Eq{PE2}. That substitution recursively expands each change in property value, $\GD z_i$, to itself be composed of a selection term and property value change term.  For each group, $i$, we now have expressions for selection within the group, $\cov(w_i,z_i)/\wbar_i$, and average property value change within the group, $\E(w_i\GD z_i)/\wbar_i$. If we write out the full expression for this last term, we obtain
\begin{equation*}
  \E(w_i\GD z_i)/\wbar_i = \sum_j w_{ij}\GD z_{ij}/\wbar_i.
\end{equation*}
In the term $\GD z_{ij}$, each labeling, $j$, may itself be a subgroup within the larger grouping represented by $i$.  The recursive nature of the Price equation allows another expansion to the characters $z_{ijk}$ for the $k$th entity in the $j$th grouping that is nested in the $i$th group, and so on.  Once again, the indexing for levels $i$, $j$, and $k$ do not have to correspond to any particular structuring, but we may choose to use a structuring if we wish.

One could analyze biological problems of group selection without using the Price equation.  Because the Price equation is a mathematical identity, there are always other ways of expressing the same thing.  However, in the 1970s, when group selection was  a very confused subject, the Price equation's recursive nature and Hamilton's development provided the foundation for subsequent understanding of the topic.  All modern conceptual insights about group selection derive from Price's recursive expansion of his abstract expression of selection.

\section*{History and alternative expressions of selection}

I have emphasized the general and abstract form of the Price equation.  That abstract form was first presented rather cryptically by \textcite{price72extension}.  In that article, Price described the recursive expansion to analyze group selection.  Apart from the recursive aspect, the more general abstract properties were hardly mentioned in \textcite{price72extension} and not developed by others until 1995.  

While I was writing my history of Price's contributions to evolutionary genetics \autocite{frank95george}, I found Price's unpublished manuscript \emph{The nature of selection} among W.~D.~Hamilton's papers.  Price's unpublished manuscript gave a very general and abstract scheme for analyzing selection in terms of set relations.  However, Price did not explicitly connect the abstract set relation scheme to the Price equation or to his earlier publications \autocite{price70selection,price72extension}.

I had \emph{The nature of selection} published posthumously as \textcite{price95the-nature}.  In my own article, I explicitly developed the general interpretation of the Price equation as the formal abstract expression of the relation between two sets \autocite{frank95george}. 

\textcite{price70selection} wrote an earlier article in which he presented a covariance selection equation that emphasized the connection to classical models of population genetics and gene frequency change. That earlier covariance form lacks the abstract set interpretation and generally has narrower scope.  Preceding Price, \textcite{robertson66a-mathematical} and 
\textcite{li67fundamental} also presented selection equations that are similar to Price's \citeyear{price70selection} covariance expression.  Robertson's covariance form itself arises from classical quantitative genetics and the breeder's equation, ultimately deriving from the foundations of quantitative genetics established by \textcite{fisher18the-correlation}.  Li's form presents a covariance type of expression for classical population genetic models of gene frequency change.

One cannot understand the current literature without a clear sense of this history.  Almost all applications of the Price equation to kin and group selection, and to other problems of evolutionary analysis, derive from either the classical expressions of quantitative genetics \autocite{robertson66a-mathematical} or classical expressions of population genetics \autocite{li67fundamental}.

In light of this history, criticisms can be confusing with regard to the ways in which the Price equation is commonly used.  For example, in applications to kin or group selection, the Price equation mainly serves to package the notation for the Robertson form of quantitative genetic analysis or the Li form of population genetic analysis.  The Price equation packaging brings no extra assumptions.  In some applications, critics may believe that the particular analysis lacks enough assumptions to attain a desired level of specificity. One can, of course, easily add more assumptions, at the expense of reduced generality.

The following sections briefly describe some alternative forms of the Price equation and the associated history.  That history helps to place criticisms of the Price equation and its applications into clearer light.

\subsection*{Quantitative genetics and the breeder's equation}

\textcite{fisher18the-correlation} established the modern theory of quantitative genetics, following the early work of Galton, Pearson, Weldon, Yule and others. The equations of selection in quantitative genetics and animal breeding arose from that foundation.  Many modern applications of the Price equation to particular problems follow this tradition of quantitative genetics.  A criticism of these Price equation applications is a criticism of the central approach of evolutionary quantitative genetics.  Such criticisms may be valid for certain applications, but they must be evaluated in the broader context of quantitative genetics theory.  This section shows the relation between quantitative genetics and a commonly applied form of the Price equation \autocite{rice04evolutionary}.

Evolutionary aspects of quantitative genetics developed from the breeder's equation
\begin{equation*}
  R = Sh^2,
\end{equation*}
in which the response to selection, $R$, equals the selection differential, $S$, multiplied by the heritability, $h^2$. The separation of selection and transmission is the key to the breeder's equation and to quantitative genetics theory. 

The covariance term of the Price equation is equivalent to the selection differential, $S$, when one interprets the meaning of \emph{fitness} and \emph{descendants} in a particular way.  Suppose that we label each potential parent in the ancestral population of size $N$ with the index, $i$.  The initial weighting of each parent in the ancestral population is $q_i=1/N$. Assign to each potential parent a weighting with respect to breeding contribution, $q'_i=q_iw_i$, with fitnesses standardized so that $\wbar=1$ and the $w_i$ are relative fitnesses.  

With this setup, ancestors are the initial population of potential parents, each weighted equally, and descendants are the same population of parents, weighted by their breeding contribution.  The character value for each individual remains unchanged between the ancestor and descendant labelings.  These assumptions lead to $\GD\zbar^*=\cov(w,z)$, the change in the average character value between the breeding population and the initial population.  That difference is defined as $S$, the selection differential.

To analyze the fraction of the selection differential transmitted to offspring, classical quantitative genetics follows \textcite{fisher18the-correlation} to separate the character value as $z=g+\Ge$, with a transmissible genetic component, $g$, and a component that is not transmitted, which we may call the environmental or unexplained component, $\Ge$.  Following standard regression theory for this sort of expression, $\bar{\Ge}=0$.  

For a parent with $z=g+\Ge$, the average character value contribution ascribed to the parent among its descendants is $z'=g$, following the idea that $g$ represents the component of the parental character that is transmitted to offspring.  If we assume that the only fluctuations of average character value in offspring are caused by the transmissible component that comes from parents, then the genetic component measured by $g$ is sufficient to explain expected offspring character values.  Thus, $\GD z = z'-z=-\Ge$, and $\E(w\GD z)=-\cov(w,\Ge)$.

Substituting into the full Price equation from \Eq{PriceEq} and assuming $\wbar=1$ so that all fitnesses are normalized
\begin{align}
  \GD\zbar &= \cov(w,z) + \E(w\GD z) \notag\\
           &= \cov(w,g) + \cov(w,\Ge) - \cov(w,\Ge) \notag\\
           &= \cov(w,g). \label{eq:secondaryTheorem}
\end{align}
The expression $\GD\zbar = \cov(w,g)$ was first emphasized by \textcite{robertson66a-mathematical}, and is sometimes called Robertson's secondary theorem of natural selection.  Robertson's expression summarizes the foundational principles of quantitative genetics, as conceived by \textcite{fisher18the-correlation} and developed over the past century \autocite{falconer96introduction,lynch98genetics,hartl06principles}.

It is commonly noted that Robertson's theorem is related to the classic breeder's equation. In particular,
\begin{equation*}
  R=\GD\zbar=\cov(w,g)=\cov(w,z)h^2=Sh^2,
\end{equation*}
where $R$ is the response to selection, $S=\cov(w,z)$ is the selection differential, and $h^2=\var(g)/\var(z)$ is a form of heritability, a measure of the transmissible genetic component.  Additional details and assumptions can be found in several articles and texts \autocite{crow76the-rate,frank97the-price,rice04evolutionary}.

\subsection*{Population genetics and the covariance expression}

\textcite{price70selection} expressed his original formulation in terms of gene frequency change and classical population genetics, rather than the abstract set relations that I have emphasized.  At that time, it seems likely that Price already had the broader, more abstract theory in hand, and was presenting the population genetics form because of its potential applications.  The article begins
\begin{quote}
This is a preliminary communication describing applications to genetical selection of a new mathematical treatment of selection in general.

Gene frequency change is the basic event in biological evolution. The following equation$\ldots$which gives frequency change under selection from one generation to the next for a single gene or for any linear function of any number of genes at any number of loci, holds for any sort of dominance or epistasis, for sexual or asexual reproduction, for random or nonrandom mating, for diploid, haploid or polyploid species, and even for imaginary species with more than two sexes$\ldots$
\end{quote}
Using my notation, Price writes the basic covariance form
\begin{equation}\label{eq:pricePopGen}
  \GD P = \cov(w,p)/\wbar = \Gb_{wp}\var(p)/\wbar.
\end{equation}
In a simple application, $p$ could be interpreted as gene frequency at a single diploid locus with two alleles. Then $P=\bar{p}$ is the gene frequency in the population, and $\Gb_{wp}$ is the regression of individual fitness on individual gene frequency, in which the individual gene frequency is either $0$, $1/2$ or $1$ for an individual with $0$, $1$ or $2$ copies of the allele of interest.   \textcite{li67fundamental} gave an identical gene frequency expression in his eqn~4.

In more general applications, one can study a $p$-score that summarizes the number of copies of various alleles present in an individual, or in whatever entities are being tracked.  In classical population genetics, the $p$-score would be, in Price's words above, ``any linear function of any number of genes at any number of loci.''  Here, \textit{linearity} means that $p$ is essentially a counting of presence versus absence of various things within the $i$th entity.  Such counting does not preclude nonlinear interactions between alleles or those things being counted with respect to phenotype, which is why Price said that the expression holds for any form of dominance or epistasis.

\textcite{hamilton70selfish} used Price's gene frequency form in his first clear derivations of the direct and the inclusive fitness models of kin selection theory.  Most early applications of the Price equation used this gene frequency interpretation.

\textcite{price70selection} emphasized that the value of \Eq{pricePopGen} arises from its benefits for qualitative reasoning rather than calculation. The necessary assumptions can be seen from the form given by Price, which is always exact, here written in my notation
\begin{equation*}
  \GD P = \cov(w,p)/\wbar + \E(w\GD p)/\wbar,
\end{equation*}
where $\GD p$ is interpreted as the change in state between parental gene frequency for the $i$th entity and the average gene frequency for the part of descendants derived from the $i$th entity.  

In practice, $\GD p=0$ usually means Mendelian segregation, no biased mutation, and no sampling biases associated with drift.  Most population genetics theory of traits such as social behavior typically make those assumptions, so that \Eq{pricePopGen} is sufficient with respect to analyzing change in gene frequency or in $p$-scores \autocite{grafen84natural}.  However, the direction of change in gene frequency or $p$-score is not sufficient to predict the direction of change in phenotype.  To associate the direction of change in $p$-score to the direction of change in phenotype, one must make the assumption that phenotype changes monotonically with $p$-score.  Such monotonicity is a strong assumption, which is not always met.  For that reason, $p$-score models sometimes buy simplicity at a rather high cost.  In other applications, monotonicity is a reasonable assumption, and the $p$-score models provide a very simple and powerful approach to understanding the direction of evolutionary change.

The costs and benefits of the $p$-score model are not particular to the Price equation.  Any analysis based on the same assumptions has the same limitations.  The Price equation provides a concise and elegant way to explore the consequences when certain simplifying assumptions can reasonably be applied to a particular problem.  

\section*{Alternative forms or interpretations of the full equation}

The full Price equation partitions total evolutionary change into components.  Many alternative partitions exist. A partition provides value if it improves conceptual clarity or eases calculation.  

Which partitions are better than others?  \emph{Better} is always partly subjective.  What may seem hard for me may appear easy to you.  Nonetheless, it would be a mistake to suggest that all differences are purely subjective.  Some forms are surely better than others for particular problems, even if \emph{better} remains hard to quantify.  As \textcite[p.~14]{russell58the-abc-of-relativity} said in another context, ``All such conventions are equally legitimate, though not all are equally convenient.'' 

Many partitions of evolutionary change include some aspect of selection and some aspect of property or transmission change. Most of those variants arise by minor rearrangements or extensions of the basic Price expression.  A few examples follow.

\subsection*{Contextual analysis}

\textcite{heisler87a-method} introduced the phrase \textit{contextual analysis} to the evolutionary literature.  Contextual analysis is a form of path analysis, which partitions causes by statistical regression models.  Path analysis has been used throughout the history of genetics \autocite{li75path}.  It is a useful approach whenever one wishes to partition variation with respect to candidate causes.  The widely used method of \textcite{lande83the-measurement} to analyze selection is a particular form of path analysis.  

\textcite{okasha06evolution} argued that contextual analysis is an alternative to the Price equation.  To develop a simple example, let us work with just the selection part of the Price equation
\begin{equation*}
	\wbar\GD\zbar = \cov(w,z).
\end{equation*}
A path  (contextual) analysis refines this expression by partitioning the causes of fitness with a regression equation.  Suppose we express fitness as depending on two predictors: the focal character that we are studying, $z$, and another character, $y$.  Then we can write fitness as
\begin{equation*}
	w = \Gb_{wz}z+\Gb_{wy}y + \Ge
\end{equation*}
in which the $\Gb$ terms are partial regressions of fitness on each character, and $\Ge$ is the unexplained residual of fitness.  Substituting into the Price equation, we get the sort of expression made popular by \textcite{lande83the-measurement}
\begin{equation*}
	\wbar\GD\zbar = \Gb_{wz}\var(z)+\Gb_{wy}\cov(y,z).
\end{equation*}

If the partitioning of fitness into causes is done in a useful way, this type of path analysis can provide significant insight.  I based my own studies of natural selection and social evolution on this approach \autocite{frank97the-price,frank98foundations}.

Authors such as \textcite{okasha06evolution} consider the partitioning of fitness into distinct causes as an alternative to the Price equation. If one thinks of the character $z$ in $\cov(w,z)$ as a complete causal explanation for fitness, then a partition into separate causes $y$ and $z$ does indeed lead to a different causal understanding of fitness.  In that regard, the Price equation and path analysis lead to different causal perspectives.  

One can find articles that use the Price equation and interpret $z$ as a lone cause of fitness \autocite<see>{okasha06evolution}.  Thus, if one equates those specific applications with the general notion of \textit{the Price equation,} then one can say that path or contextual analysis provides a significantly different perspective from the Price equation.  To me, that seems like a socially constructed notion of logic and mathematics.  If someone has applied an abstract truth in a specific way, and one can find an alternative method for the same specific application that seems more appealing, then one can say that the alternative method is superior to the general abstract truth.  

The abstract Price equation does not compel one to interpret $z$ strictly as a single cause explanation.  Rather, in the general expression, $z$ should always be interpreted as an abstract placeholder.  Path (contextual) analysis follows as a natural extension of the Price equation, in which one makes specific models of fitness expressed by regression.  It does not make sense to discuss the Price equation and path analysis as alternatives.

\subsection*{Alternative partitions of selection and transmission}

In the standard form of the Price equation, the fitness term, $w$, appears in both components 
\begin{equation*}
  \wbar\GD\zbar = \cov(w,z) + \E(w\GD z).
\end{equation*}
\textcite{frank97the-price,frank98foundations} derived an alternative expression
\begin{align}
	\GD\zbar &= \sum q_i'z_i' - \sum q_i z_i \notag\\
	                &= \sum q_i(w_i/\wbar)z_i' - \sum q_i z_i \notag\\
	                &= \sum q_i(w_i/\wbar)z_i' - \sum q_iz_i' + \sum q_iz_i' - \sum q_iz_i \notag\\
	                &= \sum q_i\left(w_i/\wbar-1\right)z_i' + \sum q_i(z_i'-z_i) \notag\\
	                &= \cov(w,z')/\wbar + \E(\GD z). \label{eq:altPart}
\end{align}
This form sometimes provides an easier method to calculate effects.  For example, the second term now expresses the average change in phenotype between parent and offspring without weighting by fitness effects.  A biased mutational process would be easy to calculate with this expression---one only needs to know about the mutation process to calculate the outcome.  The new covariance term can be partitioned into meaningful components with minor assumptions \autocite[p.~1721]{frank97the-price}, yielding
\begin{equation*}
	\cov(w,z') = \cov(w,z)\Gb_{z'z},
\end{equation*}
where $\Gb_{z'z}$ is usually interpreted as the offspring-parent regression, which is a type of heritability. Thus, we may combine selection with the heritability component of transmission into the covariance term, with the second term containing only a fitness-independent measure of change during transmission.  

\textcite{okasha06evolution} strongly favored the alternative partition for the Price equation in \Eq{altPart}, because it separates all fitness effects in the first term from a pure transmission interpretation of the second term.  In my view, there are costs and benefits for the standard Price equation expression compared with \Eq{altPart}.  One gains by having both, and using the particular form that fits a particular problem.  

For example, the term $\E(\GD z)$ is useful when one has to calculate the effects of a biased mutational process that operates independently of fitness.  Alternatively, suppose most individuals have unbiased transmission, such that $\GD z=0$, whereas very sick individuals do not reproduce but, if they were to reproduce, would have a very biased transmission process.  Then $\E(\GD z)$ differs significantly from zero, because the sick, nonreproducing individuals appear in this term equally with the reproducing population.  However, the actual transmission bias that occurs in the population would be zero, $\E(w\GD z)=0$, because all reproducing individuals have nonbiased transmission.  

Both the standard Price form and the alternative in \Eq{altPart} can be useful.  Different scenarios favor different ways of expressing problems.  I cannot understand why one would adopt an a priori position that unduly limits one's perspective.

\subsection*{Extended set mapping expression}

The Price equation's power arises from its abstraction of selection in terms of mapping relations between sets \autocite{frank95george,price95the-nature}. Although the Price equation is widely cited in the literature, almost no work has developed the set mapping formalism beyond the description given in the initial publications.  I know of only one article. 

\textcite{kerr09generalization} noted that, in the original Price formulation, every descendant must derive from one or more ancestors.  There is no natural way for novel entities to appear. In applications, new entities could arise by immigration from outside the system or, in a cultural interpretation, by de novo generation of an idea or behavior.  

\textcite{kerr09generalization} present an extended expression to handle unconnected descendants.  Their formulation depends on making explicit the connection number between each individual ancestor and each individual descendant, rather than using the fitnesses of types.  Some descendants may have zero connections. 

With an explicit description of connections, an extended Price equation follows.  The two core components of covariance for selection and expected change for transmission occur, plus a new factor to account for novel descendants unconnected to ancestors.  

The notation in \textcite{kerr09generalization} is complex, so I do not repeat it here.  Instead, I show a simplified version.  Suppose that a fraction $p$ of the descendants are unconnected to ancestors.  Then we can write the average trait value among descendants as
\begin{align*}
  \zbar' = p\sum \Ga_j \sz_j + (1-p)\sum q_i'z_i',
\end{align*}
where $\sz_j$ is the phenotype for the $j$th member of the descendant population that is unconnected to ancestors, and $\Ga_j$ is the frequency of each unconnected type, with $\sum\Ga_j=1$.  Given those definitions, we can proceed with the usual Price equation expression
\begin{align*}
  \GD\zbar &= \zbar' -\zbar \\
                  &= p\sum \Ga_j \sz_j + (1-p)\sum q_i'z_i' - (p+1-p)\sum q_iz_i\\
                  &= (1-p)\left(\sum q_i'z_i' - \sum q_iz_i\right)+ p\left(\sum \Ga_j \sz_j - \sum q_iz_i\right).
\end{align*}
Note that the term weighted by $1-p$ leads to the standard form of the Price equation, so we can write
\begin{align*}
  \GD\zbar &= (1-p)\left(\cov(w,z)+\E(w\GD z)\right)/\wbar + p\left(\sum \Ga_j \sz_j - \sum q_iz_i\right)\\
                  &= (1-p)\left(\cov(w,z)+\E(w\GD z)\right)/\wbar + p\left(\zbar^*-\zbar\right).
\end{align*}
In the component weighted by $p$, no connections exist between the descendant $\sz_j$ and a member of the ancestral population.  Thus, we have no basis to relate those terms to fitness, transmission, or property change.  \textcite{kerr09generalization} use an alternative notation that associates all entities with their number of connections, including those with zero.  The outcome is an extended set mapping theory for evolutionary change. The main concepts and the value of the approach are best explained by the application presented in the next section.

\subsection*{Gains and losses in descendants and ancestors}

\textcite{fox12analyzing} analyze changes in ecosystem function by modifying the method of \textcite{kerr09generalization}.  They measure ecosystem function by summing the functional contribution of each species present in an ecosystem.  To compare ecosystems, they consider an initial site and a second site.  When comparing ecosystems, the notion of ancestors and descendants may not make sense. Instead, one appeals to the more general set mapping relations of the Price equation.  

Assume that there is an initial site with total function $T=\sum z_i$, where $z_i$ is the function of the $i$th species.  At the initial site, there are $s$ different species, thus we may also express the total as $T=s\zbar$, where $\zbar$ is the average function per species.  At a second site, total function is $T'=\sum z_j'$, with $s'$ different species in the summation, and $T'=s'\zbar'$.  Let the number of species in common between the sites be $s_c$.  Thus, the initial site has $S=s-s_c$ unique species, and the second site has $S'=s'-s_c$ unique species.  

\textcite{fox12analyzing} write the change in total ecosystem function as
\begin{align*}
  \GD T = T' - T &= s'\zbar' - s\zbar\\
                         &= (s'-s_c)\zbar' - (s-s_c)\zbar + s_c\left(\zbar'-\zbar\right)\\
                         &= S'\zbar' - S\zbar + s_c(\GD\zbar).
\end{align*}
The term $S'\zbar'$ represents the change in function caused the gain of an average species, in which $S'$ is the number of newly added species, and $\zbar'$ is the average function per species.   Fox \& Kerr suggest that a randomly added species would be expected to function as an average species, and so interpret this term as the contribution of random species gain.  The term $S\zbar$ is interpreted similarly as random species loss with respect to the $S$ unique species in the first ecosystem not present in the second ecosystem. 

Fox \& Kerr partition the term $s_c(\GD\zbar)$ into three components of species function: deviation from the average for species gained at the second site, deviation from the average for species lost from the first site, and the changes in function for those species in common between sites.  

The point here concerns the approach rather than the theory of ecosystem function.  To analyze changes between two sets, one often benefits by an explicit decomposition of the relations between the two sets.  The original Price equation is one sort of decomposition, based on tracing the ways in which descendants derive from and change with respect to ancestors.  \textcite{fox12analyzing} extend the decomposition of change by set mapping to include specific components that make sense in the context of changes in ecosystem function.   

More work on the mathematics of set mapping and decomposition would be very valuable.  The Price equation and the extensions by Kerr, Godfrey-Smith, and Fox show the potential for thinking carefully about the abstract components of change between sets, and how to apply that abstract understanding to particular problems.
 
\subsection*{Other examples}

No clear guidelines determine what constitutes an extension to the Price equation.  From a broad perspective, many different partitions of total change have similarities, because they separate something like selection from other forces that alter the similarity between populations.  

For example, the stochastic effects of sampling and drift create a distribution of descendant phenotypes around the ancestral mean.  In the classical Price formulation, there is only the single realization of the actual descendants.  A stochastic version analyzes a collection of possible descendant sets over some probability distribution, and a mapping from the ancestor set to each possible realization of the descendant set.  

In other cases, partitions will split components more finely or add new components not in Price's formulation.  I do not have space to review every partition of total change and consider how each may be related to Price's formulation.  I list a few examples here.

\textcite{grafen99formal} and \textcite{rice08a-stochastic} developed stochastic approaches. \textcite{grafen07the-formal} based a long-term project on interpretations and extensions of the Price equation. \textcite{page02unifying} related the Price equation to various other evolutionary analyses, providing some minor extensions.  \textcite{wolf98evolutionary}, \textcite{bijma08the-joint}, and many others developed extended partitions by splitting causes with regression or similar methods such as path analysis.  Various forms of the Price equation have been applied in economic theory \autocite{andersen04population}.

\section*{Difficulties with various critiques of the Price equation}

\begin{quote}
A reliable way to make people believe in falsehoods is frequent repetition, because familiarity is not easily distinguished from truth \autocite[p.~62]{kahneman11thinking}.
\end{quote}

\noindent One must distinguish the full, exact Price equation from various derived forms used in applications.  The derived forms always make additional assumptions or express approximate relations \autocite{frank97the-price}. Each assumption increases specificity and reduces generality in relation to particular goals.

Critiques of the Price equation rarely distinguish the costs and benefits of particular assumptions in relation to particular goals.  I use van Veelen's recent series of papers as a proxy for those critiques.  That series repeats some of the common misunderstandings and adds some new ones.  Nowak recently repeated van Veelen's critique as the basis for his commentary on the Price equation \autocite{veelen05on-the-use-of-the-price,veelen10call,van-veelen11a-rule,vanveelen12group,nowak10the-evolution,nowak11supercooperators:}.

\subsection*{Dynamic sufficiency}

The Price equation describes the change in some measurement, expressed as $\GD\zbar$.  Change is calculated with respect to particular mapping relations between ancestor and descendant populations.  We can think of the mappings and the beginning value of $\zbar$ as the initial conditions or inputs, and $\GD\zbar$ as the output.  

The output, $\zbar'=\zbar+\GD\zbar$, does not provide enough information to iterate the calculation of change in order to get another value of $\GD\zbar$ starting with $\zbar'$.  We would also need the mapping relations between the new descendant population and its subsequent descendants.  That information is not part of the initial input.  Thus, we cannot study the dynamics of change over time without additional information.  

This limitation with regard to repeated iteration is called a lack of \textit{dynamic sufficiency} \autocite{lewontin74the-genetic}. Confusion about the nature of dynamic sufficiency in relation to the Price equation has been common in the literature.  In \textcite[pp.~378--379]{frank95george}, I wrote
\begin{quote}
It is not true, however, that dynamic sufficiency is a property that can be ascribed to the Price Equation---this equation is simply a mathematical tautology for the relationship among certain quantities of populations. Instead, dynamic sufficiency is a property of the assumptions and information provided in a particular problem, or added by additional assumptions contained within numerical techniques such as diffusion analysis or applied quantitative genetics. $\ldots$ What problems can the Price equation solve that cannot be solved by other
methods? The answer is, of course, none, because the Price Equation is derived from, and is no more than, a set of notational conventions. It is a mathematical tautology.
\end{quote}
I showed how the Price equation helps to define the necessary conditions for dynamic sufficiency. Once again, the Price equation proves valuable for clarifying the abstract structure of evolutionary analysis.  

Compare my statement to \textcite{vanveelen12group}
\begin{quote}
Dynamic insufficiency is regularly mentioned as a drawback of the Price equation (see for example Frank, 1995; Rice, 2004). We think that this is not an entirely accurate description of the problem. We would like to argue that the perception of dynamic insufficiency is a symptom of the fundamental problem with the Price equation, and not just a drawback of an otherwise fine way to describe evolution. To begin with, it is important to realize that the Price equation itself, by its very nature, cannot be dynamically sufficient or insufficient. The Price equation is just an identity. If we are given a list of numbers that represent a transition from one generation to the next, then we can fill in those numbers in both the right and the left hand side of the Price equation. The fact that it is an identity guarantees that the numbers that appear on both sides of the equality sign are the same. There is nothing dynamically sufficient or insufficient about that (this point is also made by Gardner et al., 2007, p. 209). A model, on the other hand, can be dynamically sufficient or insufficient.
\end{quote}

This quote from \textcite{vanveelen12group} demonstrates an interesting approach to scholarship.  They first cite Frank as stating that dynamic insufficiency is a drawback of the Price equation.  They then disagree with that point of view, and present as their own interpretation an argument that is nearly identical in concept and phrasing to my own statement in the very paper that they cited as the foundation for their disagreement.  

In this case, I think it is important to clarify the concepts and history, because influential and widely cited authors, such as Nowak, are using van Veelen's articles as the basis for their own critiques of the Price equation and approaches to fundamental issues of evolutionary analysis.  

With regard to dynamics, any analysis achieves the same dynamic status given the same underlying assumptions.  The Price equation, when used with the same underlying assumptions as population genetics, has the same attributes of dynamic sufficiency as population genetics. 

\subsection*{Interpretation of covariance}

\textcite{vanveelen12group} claim that 
\begin{quote}
Maybe the most unfortunate thing about the Price equation is that the term on the right hand side is denoted as a covariance, even though it is not. The equation thereby turns into something that can easily set us off in the wrong direction, because it now resembles equations as they feature in other sciences, where probabilistic models are used that do use actual covariances. 
\end{quote}

One can see the covariance expression in the standard form of the Price equation given in \Eq{PriceEq}. In the Price equation, the covariance is measured with respect to the total \emph{population}, in other words, it expresses the association over all members of the population.  In many statistical applications, one only has data on a subset of the full population, that subset forming a \emph{sample}.  It is important to distinguish between population measures and sample measures, because they refer to different things.  

\textcite[p.~485]{price72extension} made clear that his equation is about total change in entire populations, so the covariance is interpreted as a population measure
\begin{quote}
[W]e will be concerned with population functions and make no use of sample functions, hence we will not observe notational conventions for distinguishing population and sample variables and functions.
\end{quote}

In additional to population and sample measures, covariance also arises in mathematical models of process.  Suppose, for example, that I develop a model in which random processes influence fitness and random processes influence phenotype.  If the random fluctuations in fitness and the random fluctuations in phenotype are associated, the random variables of fitness and phenotype would covary.  All of these different interpretations of covariance are legitimate, they simply reflect different situations.

\section*{Discussion}

In \textcite{frank95george}, I wrote: ``What problems can the Price equation solve that cannot be solved by other methods? The answer is, of course, none, because the Price Equation is derived from, and is no more than, a set of notational conventions. It is a mathematical tautology.''

\textcite{nowak11supercooperators:} and \textcite{vanveelen12group} emphasize the same point in their critique of the Price equation, although they present the argument as a novel insight without attribution.  Given that the Price equation is a set of notational conventions, it cannot uniquely specify any predictions or insights.  A particular set of assumptions leads to the same predictions, no matter what notational conventions one uses.  The Price equation is a tool that sometimes helps in analysis or in seeing general connections between apparently disparate ideas.  For many problems, the Price equation provides no value, because it is the wrong tool for the job.

If the Price equation is just an equivalence, or tautology, then why am I enthusiastic about it?  Mathematics is, in its essence, about equivalences, as expressed beautifully in the epigraph from Mazur.  Not all equivalences are interesting or useful, but some are, just as not all mathematical expressions are interesting or useful, but some are.  

That leads us to the question of how we might know whether the Price equation is truly useful or a \emph{mere} identity?  It is not always easy to say exactly what makes an abstract mathematical equivalence interesting or useful.  However, given the controversy over the Price equation, we should try.  Because there is no single answer, or even a truly unique and unambiguous question, the problem remains open. I list a few potential factors.

``[A] good notation has a subtlety and suggestiveness which at times make it seem almost like a live teacher'' \autocite[pp.~17--18]{russell22introduction}. Much of creativity and understanding comes from seeing previously hidden associations. The tools and forms of expression that we use play a strong role in suggesting connections and are inseparable from cognition \autocite{kahneman11thinking}.  Equivalences and alternative notations are important. 

The various forms of the covariance component from the Price equation given in \Eq{altCov} show the equivalence of the statistical, geometrical and informational expressions for natural selection. The recursive form of the full Price equation provides the foundation for all modern studies of group selection and multilevel analysis. The Price equation helped in discovering those various connections, although there are many other ways in which to derive the same relations. 

\textcite{hardy67a-mathematicians} also emphasized the importance of seeing new connections between apparently disparate ideas:
\begin{quote}
We may say, roughly, that a mathematical idea is `significant' if it can be connected, in a natural and illuminating way, with a large
complex of other mathematical ideas. Thus a serious mathematical theorem, a theorem which connects significant ideas, is likely to lead to important advances in mathematics itself and even in other sciences.
\end{quote}

What sort of connections?  One type concerns the invariances discovered or illuminated by the Price equation.  I discussed some of those invariances in an earlier section, particularly the information theory interpretation of natural selection through the measure of Fisher information \autocite{frank09natural}.  Fisher's fundamental theorem of natural selection is a similar sort of invariance \autocite{frank12wrights}. Kin selection theory derives much of its power by identifying an invariant informational quantity sufficient to unify a wide variety of seemingly disparate processes \autocite[Chapter 6]{frank98foundations}. The interpretation of kin selection as an informational invariance has not been fully developed and remains an open problem.

Invariances provide the foundation of scientific understanding: ``It is only slightly overstating the case to say that physics is the study of symmetry'' \autocite{anderson72more}. \textit{Invariance} and \textit{symmetry} mean the same thing \autocite{weyl83symmetry}.  \textcite{feynman67the-character} emphasized that invariance is \textit{The Character of Physical Law.}  The commonly observed patterns of probability can be unified by the study of invariance and its association to measurement \autocite{frank10measurement,frank11a-simple}. There has been little effort in biology to pursue similar understanding of invariance and measurement \autocite{frank11measurement,houle11measurement}. 

Price argued for the great value of abstraction, in the sense of the epigraph from Mazur.  In \textcite{price95the-nature}
\begin{quote}
[D]espite the pervading importance of selection in science and life, there has been no abstraction and generalization from genetical selection to obtain a general selection theory and general selection mathematics. Instead, particular selection problems are treated in ways appropriate to particular fields of science. Thus one might say that `selection theory' is a theory waiting to be born---much as communication theory was 50 years ago. Probably the main lack that has been holding back any development of a general selection theory is lack of a clear concept of the general nature or meaning of `selection'.
\end{quote}

This article has been about the Price equation in relation to its abstract properties and its connections to various topics, such as information or fundamental invariances.  Some readers may feel that those aspects of abstraction and invariance are nice, but far from daily work in biology.  What of the many applications of the Price equation to kin or group selection?  Do those applications hold up?  How much value has been added?  

Because the Price equation is a tool, one can always arrive at the same result by other methods.  How well the Price equation works depends partly on the goal and partly on the fit of the tool to the problem.  There is inevitably a strongly subjective aspect to deciding about how well a tool works.  Nonetheless, hammers truly are good for nails and bad for screws. For valuing tools, there is a certain component that should be open to agreement.  For example, the \textcite{robertson66a-mathematical} form of the Price equation is widely regarded as the foundational method for analyzing models of evolutionary quantitative genetics.  However, not all problems in quantitative genetics are best studied with the Robertson-Price equation.  And not all problems in social evolution benefit from a Price equation approach.

The Price equation or descendant methods have led to many useful models for kin selection \autocite{frank98foundations}.  The most powerful follow a path analysis decomposition of causes or use a simple maximization method to analyze easily what would otherwise have been difficult.  I will return to those applications in subsequent articles.
 
\section*{Acknowledgments}

I thank R.~M.~Bush and W.~J.~Ewens for helpful comments. My research is supported by National Science Foundation grant EF-0822399, National Institute of General Medical Sciences MIDAS Program grant U01-GM-76499, and a grant from the James S.~McDonnell Foundation.  

\bibliography{main}

\begin{thebibliography}{77}%
\makeatletter
\providecommand \@ifxundefined [1]{%
 \@ifx{#1\undefined}
}%
\providecommand \@ifnum [1]{%
 \ifnum #1\expandafter \@firstoftwo
 \else \expandafter \@secondoftwo
 \fi
}%
\providecommand \@ifx [1]{%
 \ifx #1\expandafter \@firstoftwo
 \else \expandafter \@secondoftwo
 \fi
}%
\providecommand \natexlab [1]{#1}%
\providecommand \enquote  [1]{``#1''}%
\providecommand \bibnamefont  [1]{#1}%
\providecommand \bibfnamefont [1]{#1}%
\providecommand \citenamefont [1]{#1}%
\providecommand \href@noop [0]{\@secondoftwo}%
\providecommand \href [0]{\begingroup \@sanitize@url \@href}%
\providecommand \@href[1]{\@@startlink{#1}\@@href}%
\providecommand \@@href[1]{\endgroup#1\@@endlink}%
\providecommand \@sanitize@url [0]{\catcode `\\12\catcode `\$12\catcode
  `\&12\catcode `\#12\catcode `\^12\catcode `\_12\catcode `\%12\relax}%
\providecommand \@@startlink[1]{}%
\providecommand \@@endlink[0]{}%
\providecommand \url  [0]{\begingroup\@sanitize@url \@url }%
\providecommand \@url [1]{\endgroup\@href {#1}{\urlprefix }}%
\providecommand \urlprefix  [0]{URL }%
\providecommand \Eprint [0]{\href }%
\providecommand \doibase [0]{http://dx.doi.org/}%
\providecommand \selectlanguage [0]{\@gobble}%
\providecommand \bibinfo  [0]{\@secondoftwo}%
\providecommand \bibfield  [0]{\@secondoftwo}%
\providecommand \translation [1]{[#1]}%
\providecommand \BibitemOpen [0]{}%
\providecommand \bibitemStop [0]{}%
\providecommand \bibitemNoStop [0]{.\EOS\space}%
\providecommand \EOS [0]{\spacefactor3000\relax}%
\providecommand \BibitemShut  [1]{\csname bibitem#1\endcsname}%
\let\auto@bib@innerbib\@empty
%</preamble>
\bibitem [{\citenamefont {Mazur}(2008)}]{mazur08when}%
  \BibitemOpen
  \bibfield  {author} {\bibinfo {author} {\bibfnamefont {B.}~\bibnamefont
  {Mazur}},\ }\bibfield  {title} {\enquote {\bibinfo {title} {When is one thing
  equal to some other thing?}}\ }in\ \href@noop {} {\emph {\bibinfo {booktitle}
  {{Proof and Other Dilemmas: Mathematics and Philosophy}}}},\ \bibinfo
  {editor} {edited by\ \bibinfo {editor} {\bibfnamefont {B.}~\bibnamefont
  {Gold}}\ and\ \bibinfo {editor} {\bibfnamefont {R.~A.}\ \bibnamefont
  {Simons}}}\ (\bibinfo  {publisher} {Mathematical Association of America},\
  \bibinfo {address} {Washington, D.C.},\ \bibinfo {year} {2008})\ pp.\
  \bibinfo {pages} {221--241}\BibitemShut {NoStop}%
\bibitem [{\citenamefont {Boulding}(1956)}]{boulding56general}%
  \BibitemOpen
  \bibfield  {author} {\bibinfo {author} {\bibfnamefont {K.~E.}\ \bibnamefont
  {Boulding}},\ }\bibfield  {title} {\enquote {\bibinfo {title} {General
  systems theory: the skeleton of science},}\ }\href@noop {} {\bibfield
  {journal} {\bibinfo  {journal} {Management Science}\ }\textbf {\bibinfo
  {volume} {2}},\ \bibinfo {pages} {197--208} (\bibinfo {year}
  {1956})}\BibitemShut {NoStop}%
\bibitem [{\citenamefont {Price}(1970)}]{price70selection}%
  \BibitemOpen
  \bibfield  {author} {\bibinfo {author} {\bibfnamefont {G.~R.}\ \bibnamefont
  {Price}},\ }\bibfield  {title} {\enquote {\bibinfo {title} {Selection and
  covariance},}\ }\href@noop {} {\bibfield  {journal} {\bibinfo  {journal}
  {Nature}\ }\textbf {\bibinfo {volume} {227}},\ \bibinfo {pages} {520--521}
  (\bibinfo {year} {1970})}\BibitemShut {NoStop}%
\bibitem [{\citenamefont {Price}(1972{\natexlab{a}})}]{price72extension}%
  \BibitemOpen
  \bibfield  {author} {\bibinfo {author} {\bibfnamefont {G.~R.}\ \bibnamefont
  {Price}},\ }\bibfield  {title} {\enquote {\bibinfo {title} {Extension of
  covariance selection mathematics},}\ }\href@noop {} {\bibfield  {journal}
  {\bibinfo  {journal} {Annals of Human Genetics}\ }\textbf {\bibinfo {volume}
  {35}},\ \bibinfo {pages} {485--490} (\bibinfo {year}
  {1972}{\natexlab{a}})}\BibitemShut {NoStop}%
\bibitem [{\citenamefont {Nowak}\ and\ \citenamefont
  {Highfield}(2011)}]{nowak11supercooperators:}%
  \BibitemOpen
  \bibfield  {author} {\bibinfo {author} {\bibfnamefont {M.~A.}\ \bibnamefont
  {Nowak}}\ and\ \bibinfo {author} {\bibfnamefont {R.}~\bibnamefont
  {Highfield}},\ }\href@noop {} {\emph {\bibinfo {title} {{SuperCooperators:
  Altruism, Evolution, and Why We Need Each Other to Succeed}}}}\ (\bibinfo
  {publisher} {Free Press},\ \bibinfo {address} {New York},\ \bibinfo {year}
  {2011})\BibitemShut {NoStop}%
\bibitem [{\citenamefont {van Veelen}\ \emph {et~al.}(2012)\citenamefont {van
  Veelen}, \citenamefont {Garc{\'\i}a}, \citenamefont {Sabelis},\ and\
  \citenamefont {Egas}}]{vanveelen12group}%
  \BibitemOpen
  \bibfield  {author} {\bibinfo {author} {\bibfnamefont {M.}~\bibnamefont {van
  Veelen}}, \bibinfo {author} {\bibfnamefont {J.}~\bibnamefont {Garc{\'\i}a}},
  \bibinfo {author} {\bibfnamefont {M.}~\bibnamefont {Sabelis}}, \ and\
  \bibinfo {author} {\bibfnamefont {M.}~\bibnamefont {Egas}},\ }\bibfield
  {title} {\enquote {\bibinfo {title} {Group selection and inclusive fitness
  are \emph{not} equivalent; the {P}rice equation vs.\ models and
  statistics},}\ }\href@noop {} {\bibfield  {journal} {\bibinfo  {journal}
  {Journal of Theoretical Biology}\ }\textbf {\bibinfo {volume} {299}},\
  \bibinfo {pages} {64--80} (\bibinfo {year} {2012})}\BibitemShut {NoStop}%
\bibitem [{\citenamefont {Frank}(1995)}]{frank95george}%
  \BibitemOpen
  \bibfield  {author} {\bibinfo {author} {\bibfnamefont {S.~A.}\ \bibnamefont
  {Frank}},\ }\bibfield  {title} {\enquote {\bibinfo {title} {George
  {{P}}rice's contributions to evolutionary genetics},}\ }\href@noop {}
  {\bibfield  {journal} {\bibinfo  {journal} {Journal of Theoretical Biology}\
  }\textbf {\bibinfo {volume} {175}},\ \bibinfo {pages} {373--388} (\bibinfo
  {year} {1995})}\BibitemShut {NoStop}%
\bibitem [{\citenamefont {Kerr}\ and\ \citenamefont
  {Godfrey-Smith}(2009)}]{kerr09generalization}%
  \BibitemOpen
  \bibfield  {author} {\bibinfo {author} {\bibfnamefont {B.}~\bibnamefont
  {Kerr}}\ and\ \bibinfo {author} {\bibfnamefont {P.}~\bibnamefont
  {Godfrey-Smith}},\ }\bibfield  {title} {\enquote {\bibinfo {title}
  {Generalization of the {P}rice equation for evolutionary change},}\
  }\href@noop {} {\bibfield  {journal} {\bibinfo  {journal} {Evolution}\
  }\textbf {\bibinfo {volume} {63}},\ \bibinfo {pages} {531--536} (\bibinfo
  {year} {2009})}\BibitemShut {NoStop}%
\bibitem [{\citenamefont {Hamilton}(1975)}]{hamilton75innate}%
  \BibitemOpen
  \bibfield  {author} {\bibinfo {author} {\bibfnamefont {W.~D.}\ \bibnamefont
  {Hamilton}},\ }\bibfield  {title} {\enquote {\bibinfo {title} {Innate social
  aptitudes of man: an approach from evolutionary genetics},}\ }in\ \href@noop
  {} {\emph {\bibinfo {booktitle} {{Biosocial Anthropology}}}},\ \bibinfo
  {editor} {edited by\ \bibinfo {editor} {\bibfnamefont {R.}~\bibnamefont
  {Fox}}}\ (\bibinfo  {publisher} {Wiley},\ \bibinfo {address} {New York},\
  \bibinfo {year} {1975})\ pp.\ \bibinfo {pages} {133--155}\BibitemShut
  {NoStop}%
\bibitem [{\citenamefont {Frank}(1997)}]{frank97the-price}%
  \BibitemOpen
  \bibfield  {author} {\bibinfo {author} {\bibfnamefont {S.~A.}\ \bibnamefont
  {Frank}},\ }\bibfield  {title} {\enquote {\bibinfo {title} {The {{P}}rice
  equation, {{F}}isher's fundamental theorem, kin selection, and causal
  analysis},}\ }\href@noop {} {\bibfield  {journal} {\bibinfo  {journal}
  {Evolution}\ }\textbf {\bibinfo {volume} {51}},\ \bibinfo {pages}
  {1712--1729} (\bibinfo {year} {1997})}\BibitemShut {NoStop}%
\bibitem [{\citenamefont {Grafen}(2002)}]{grafen02a-first}%
  \BibitemOpen
  \bibfield  {author} {\bibinfo {author} {\bibfnamefont {A.}~\bibnamefont
  {Grafen}},\ }\bibfield  {title} {\enquote {\bibinfo {title} {A first formal
  link between the {P}rice equation and an optimization program},}\ }\href@noop
  {} {\bibfield  {journal} {\bibinfo  {journal} {Journal of Theoretical
  Biology}\ }\textbf {\bibinfo {volume} {217}},\ \bibinfo {pages} {75--91}
  (\bibinfo {year} {2002})}\BibitemShut {NoStop}%
\bibitem [{\citenamefont {Page}\ and\ \citenamefont
  {Nowak}(2002)}]{page02unifying}%
  \BibitemOpen
  \bibfield  {author} {\bibinfo {author} {\bibfnamefont {K.~M.}\ \bibnamefont
  {Page}}\ and\ \bibinfo {author} {\bibfnamefont {M.~A.}\ \bibnamefont
  {Nowak}},\ }\bibfield  {title} {\enquote {\bibinfo {title} {Unifying
  evolutionary dynamics},}\ }\href@noop {} {\bibfield  {journal} {\bibinfo
  {journal} {Journal of Theoretical Biology}\ }\textbf {\bibinfo {volume}
  {219}},\ \bibinfo {pages} {93--98} (\bibinfo {year} {2002})}\BibitemShut
  {NoStop}%
\bibitem [{\citenamefont {Andersen}(2004)}]{andersen04population}%
  \BibitemOpen
  \bibfield  {author} {\bibinfo {author} {\bibfnamefont {E.~S.}\ \bibnamefont
  {Andersen}},\ }\bibfield  {title} {\enquote {\bibinfo {title} {Population
  thinking, {P}rice's equation and the analysis of economic evolution},}\
  }\href@noop {} {\bibfield  {journal} {\bibinfo  {journal} {Evolutionary and
  Institutional Economics Review}\ }\textbf {\bibinfo {volume} {1}},\ \bibinfo
  {pages} {127--148} (\bibinfo {year} {2004})}\BibitemShut {NoStop}%
\bibitem [{\citenamefont {Rice}(2004)}]{rice04evolutionary}%
  \BibitemOpen
  \bibfield  {author} {\bibinfo {author} {\bibfnamefont {S.~H.}\ \bibnamefont
  {Rice}},\ }\href@noop {} {\emph {\bibinfo {title} {{Evolutionary Theory:
  Mathematical and Conceptual Foundations}}}}\ (\bibinfo  {publisher} {Sinauer
  Associates},\ \bibinfo {address} {Sunderland, MA},\ \bibinfo {year}
  {2004})\BibitemShut {NoStop}%
\bibitem [{\citenamefont {Okasha}(2006)}]{okasha06evolution}%
  \BibitemOpen
  \bibfield  {author} {\bibinfo {author} {\bibfnamefont {S.}~\bibnamefont
  {Okasha}},\ }\href@noop {} {\emph {\bibinfo {title} {{Evolution and the
  Levels of Selection}}}}\ (\bibinfo  {publisher} {Oxford University Press},\
  \bibinfo {address} {New York},\ \bibinfo {year} {2006})\BibitemShut {NoStop}%
\bibitem [{\citenamefont {Gardner}(2008)}]{gardner08the-price}%
  \BibitemOpen
  \bibfield  {author} {\bibinfo {author} {\bibfnamefont {A.}~\bibnamefont
  {Gardner}},\ }\bibfield  {title} {\enquote {\bibinfo {title} {The {P}rice
  equation},}\ }\href@noop {} {\bibfield  {journal} {\bibinfo  {journal}
  {Current Biology}\ }\textbf {\bibinfo {volume} {18}},\ \bibinfo {pages}
  {R198--R202} (\bibinfo {year} {2008})}\BibitemShut {NoStop}%
\bibitem [{\citenamefont {Hamilton}(1970)}]{hamilton70selfish}%
  \BibitemOpen
  \bibfield  {author} {\bibinfo {author} {\bibfnamefont {W.~D.}\ \bibnamefont
  {Hamilton}},\ }\bibfield  {title} {\enquote {\bibinfo {title} {Selfish and
  spiteful behaviour in an evolutionary model},}\ }\href@noop {} {\bibfield
  {journal} {\bibinfo  {journal} {Nature}\ }\textbf {\bibinfo {volume} {228}},\
  \bibinfo {pages} {1218--1220} (\bibinfo {year} {1970})}\BibitemShut {NoStop}%
\bibitem [{\citenamefont {Wade}(1985)}]{wade85soft}%
  \BibitemOpen
  \bibfield  {author} {\bibinfo {author} {\bibfnamefont {M.~J.}\ \bibnamefont
  {Wade}},\ }\bibfield  {title} {\enquote {\bibinfo {title} {Soft selection,
  hard selection, kin selection, and group selection},}\ }\href@noop {}
  {\bibfield  {journal} {\bibinfo  {journal} {American Naturalist}\ }\textbf
  {\bibinfo {volume} {125}},\ \bibinfo {pages} {61--73} (\bibinfo {year}
  {1985})}\BibitemShut {NoStop}%
\bibitem [{\citenamefont {Frank}\ and\ \citenamefont
  {Slatkin}(1990)}]{frank90the-distribution}%
  \BibitemOpen
  \bibfield  {author} {\bibinfo {author} {\bibfnamefont {S.~A.}\ \bibnamefont
  {Frank}}\ and\ \bibinfo {author} {\bibfnamefont {M.}~\bibnamefont
  {Slatkin}},\ }\bibfield  {title} {\enquote {\bibinfo {title} {The
  distribution of allelic effects under mutation and selection},}\ }\href@noop
  {} {\bibfield  {journal} {\bibinfo  {journal} {Genetical Research}\ }\textbf
  {\bibinfo {volume} {55}},\ \bibinfo {pages} {111--117} (\bibinfo {year}
  {1990})}\BibitemShut {NoStop}%
\bibitem [{\citenamefont {Queller}(1992{\natexlab{a}})}]{queller92a-general}%
  \BibitemOpen
  \bibfield  {author} {\bibinfo {author} {\bibfnamefont {D.~C.}\ \bibnamefont
  {Queller}},\ }\bibfield  {title} {\enquote {\bibinfo {title} {A general model
  for kin selection},}\ }\href@noop {} {\bibfield  {journal} {\bibinfo
  {journal} {Evolution}\ }\textbf {\bibinfo {volume} {46}},\ \bibinfo {pages}
  {376--380} (\bibinfo {year} {1992}{\natexlab{a}})}\BibitemShut {NoStop}%
\bibitem [{\citenamefont
  {Queller}(1992{\natexlab{b}})}]{queller92quantitative}%
  \BibitemOpen
  \bibfield  {author} {\bibinfo {author} {\bibfnamefont {D.~C.}\ \bibnamefont
  {Queller}},\ }\bibfield  {title} {\enquote {\bibinfo {title} {Quantitative
  genetics, inclusive fitness, and group selection},}\ }\href@noop {}
  {\bibfield  {journal} {\bibinfo  {journal} {American Naturalist}\ }\textbf
  {\bibinfo {volume} {139}},\ \bibinfo {pages} {540--558} (\bibinfo {year}
  {1992}{\natexlab{b}})}\BibitemShut {NoStop}%
\bibitem [{\citenamefont {Michod}(1997{\natexlab{a}})}]{michod97evolution}%
  \BibitemOpen
  \bibfield  {author} {\bibinfo {author} {\bibfnamefont {R.~E.}\ \bibnamefont
  {Michod}},\ }\bibfield  {title} {\enquote {\bibinfo {title} {Evolution of the
  individual},}\ }\href@noop {} {\bibfield  {journal} {\bibinfo  {journal} {The
  American Naturalist}\ }\textbf {\bibinfo {volume} {150}},\ \bibinfo {pages}
  {S5--S21} (\bibinfo {year} {1997}{\natexlab{a}})}\BibitemShut {NoStop}%
\bibitem [{\citenamefont {Michod}(1997{\natexlab{b}})}]{michod97cooperation}%
  \BibitemOpen
  \bibfield  {author} {\bibinfo {author} {\bibfnamefont {R.~E.}\ \bibnamefont
  {Michod}},\ }\bibfield  {title} {\enquote {\bibinfo {title} {Cooperation and
  conflict in the evolution of individuality. {I}. {M}ultilevel selection of
  the organism},}\ }\href@noop {} {\bibfield  {journal} {\bibinfo  {journal}
  {American Naturalist}\ }\textbf {\bibinfo {volume} {149}},\ \bibinfo {pages}
  {607--645} (\bibinfo {year} {1997}{\natexlab{b}})}\BibitemShut {NoStop}%
\bibitem [{\citenamefont {Frank}(1998)}]{frank98foundations}%
  \BibitemOpen
  \bibfield  {author} {\bibinfo {author} {\bibfnamefont {S.~A.}\ \bibnamefont
  {Frank}},\ }\href@noop {} {\emph {\bibinfo {title} {Foundations of {S}ocial
  {E}volution}}}\ (\bibinfo  {publisher} {Princeton University Press},\
  \bibinfo {address} {Princeton, New Jersey},\ \bibinfo {year}
  {1998})\BibitemShut {NoStop}%
\bibitem [{\citenamefont {Fox}(2006)}]{fox06using}%
  \BibitemOpen
  \bibfield  {author} {\bibinfo {author} {\bibfnamefont {J.~W.}\ \bibnamefont
  {Fox}},\ }\bibfield  {title} {\enquote {\bibinfo {title} {Using the {P}rice
  {E}quation to partition the effects of biodiversity loss on ecosystem
  function},}\ }\href@noop {} {\bibfield  {journal} {\bibinfo  {journal}
  {Ecology}\ }\textbf {\bibinfo {volume} {87}},\ \bibinfo {pages} {2687--2696}
  (\bibinfo {year} {2006})}\BibitemShut {NoStop}%
\bibitem [{\citenamefont {Day}\ and\ \citenamefont
  {Gandon}(2006)}]{day06insights}%
  \BibitemOpen
  \bibfield  {author} {\bibinfo {author} {\bibfnamefont {T.}~\bibnamefont
  {Day}}\ and\ \bibinfo {author} {\bibfnamefont {S.}~\bibnamefont {Gandon}},\
  }\bibfield  {title} {\enquote {\bibinfo {title} {Insights from {P}rice's
  equation into evolutionary epidemiology},}\ }in\ \href@noop {} {\emph
  {\bibinfo {booktitle} {{Disease Evolution: Models, Concepts, and Data
  Analysis}}}},\ \bibinfo {editor} {edited by\ \bibinfo {editor} {\bibfnamefont
  {Z.}~\bibnamefont {Feng}}, \bibinfo {editor} {\bibfnamefont {U.}~\bibnamefont
  {Dieckmann}}, \ and\ \bibinfo {editor} {\bibfnamefont {S.}~\bibnamefont
  {Levin}}}\ (\bibinfo  {publisher} {American Mathematical Society},\ \bibinfo
  {address} {Washington, D.C.},\ \bibinfo {year} {2006})\ pp.\ \bibinfo {pages}
  {23--44}\BibitemShut {NoStop}%
\bibitem [{\citenamefont {Grafen}(2007)}]{grafen07the-formal}%
  \BibitemOpen
  \bibfield  {author} {\bibinfo {author} {\bibfnamefont {A.}~\bibnamefont
  {Grafen}},\ }\bibfield  {title} {\enquote {\bibinfo {title} {The formal
  {D}arwinism project: a mid-term report},}\ }\href@noop {} {\bibfield
  {journal} {\bibinfo  {journal} {Journal of Evolutionary Biology}\ }\textbf
  {\bibinfo {volume} {20}},\ \bibinfo {pages} {1243--1254} (\bibinfo {year}
  {2007})}\BibitemShut {NoStop}%
\bibitem [{\citenamefont {Alizon}(2009)}]{alizon09the-price}%
  \BibitemOpen
  \bibfield  {author} {\bibinfo {author} {\bibfnamefont {S.}~\bibnamefont
  {Alizon}},\ }\bibfield  {title} {\enquote {\bibinfo {title} {The {P}rice
  equation framework to study disease within-host evolution},}\ }\href@noop {}
  {\bibfield  {journal} {\bibinfo  {journal} {Journal of Evolutionary Biology}\
  }\textbf {\bibinfo {volume} {22}},\ \bibinfo {pages} {1123--1132} (\bibinfo
  {year} {2009})}\BibitemShut {NoStop}%
\bibitem [{\citenamefont {Robertson}(1966)}]{robertson66a-mathematical}%
  \BibitemOpen
  \bibfield  {author} {\bibinfo {author} {\bibfnamefont {A.}~\bibnamefont
  {Robertson}},\ }\bibfield  {title} {\enquote {\bibinfo {title} {A
  mathematical model of the culling process in dairy cattle},}\ }\href@noop {}
  {\bibfield  {journal} {\bibinfo  {journal} {Animal Production}\ }\textbf
  {\bibinfo {volume} {8}},\ \bibinfo {pages} {95--108} (\bibinfo {year}
  {1966})}\BibitemShut {NoStop}%
\bibitem [{\citenamefont {Falconer}\ and\ \citenamefont
  {Mackay}(1996)}]{falconer96introduction}%
  \BibitemOpen
  \bibfield  {author} {\bibinfo {author} {\bibfnamefont {D.~S.}\ \bibnamefont
  {Falconer}}\ and\ \bibinfo {author} {\bibfnamefont {T.~F.~C.}\ \bibnamefont
  {Mackay}},\ }\href@noop {} {\emph {\bibinfo {title} {Introduction to
  {Q}uantitative {G}enetics}}},\ \bibinfo {edition} {4th}\ ed.\ (\bibinfo
  {publisher} {Longman},\ \bibinfo {address} {Essex, England},\ \bibinfo {year}
  {1996})\BibitemShut {NoStop}%
\bibitem [{\citenamefont {Charlesworth}\ and\ \citenamefont
  {Charlesworth}(2010)}]{charlesworth10elements}%
  \BibitemOpen
  \bibfield  {author} {\bibinfo {author} {\bibfnamefont {B.}~\bibnamefont
  {Charlesworth}}\ and\ \bibinfo {author} {\bibfnamefont {D.}~\bibnamefont
  {Charlesworth}},\ }\href@noop {} {\emph {\bibinfo {title} {{Elements of
  Evolutionary Genetics}}}}\ (\bibinfo  {publisher} {Roberts \& Company},\
  \bibinfo {address} {Greenwood Village, CO},\ \bibinfo {year}
  {2010})\BibitemShut {NoStop}%
\bibitem [{\citenamefont {Lande}\ and\ \citenamefont
  {Arnold}(1983)}]{lande83the-measurement}%
  \BibitemOpen
  \bibfield  {author} {\bibinfo {author} {\bibfnamefont {R.}~\bibnamefont
  {Lande}}\ and\ \bibinfo {author} {\bibfnamefont {S.~J.}\ \bibnamefont
  {Arnold}},\ }\bibfield  {title} {\enquote {\bibinfo {title} {The measurement
  of selection on correlated characters},}\ }\href@noop {} {\bibfield
  {journal} {\bibinfo  {journal} {Evolution}\ }\textbf {\bibinfo {volume}
  {37}},\ \bibinfo {pages} {1212--1226} (\bibinfo {year} {1983})}\BibitemShut
  {NoStop}%
\bibitem [{\citenamefont {Harman}(2010)}]{harman10the-price}%
  \BibitemOpen
  \bibfield  {author} {\bibinfo {author} {\bibfnamefont {O.~S.}\ \bibnamefont
  {Harman}},\ }\href@noop {} {\emph {\bibinfo {title} {{The Price of
  Altruism}}}}\ (\bibinfo  {publisher} {Norton},\ \bibinfo {address} {New
  York},\ \bibinfo {year} {2010})\BibitemShut {NoStop}%
\bibitem [{\citenamefont {Schwartz}(2000)}]{schwartz00death}%
  \BibitemOpen
  \bibfield  {author} {\bibinfo {author} {\bibfnamefont {J.}~\bibnamefont
  {Schwartz}},\ }\bibfield  {title} {\enquote {\bibinfo {title} {Death of an
  altruist},}\ }\href@noop {} {\bibfield  {journal} {\bibinfo  {journal}
  {Lingua Franca}\ }\textbf {\bibinfo {volume} {10}},\ \bibinfo {pages}
  {51--61} (\bibinfo {year} {2000})}\BibitemShut {NoStop}%
\bibitem [{\citenamefont {Price}(1995)}]{price95the-nature}%
  \BibitemOpen
  \bibfield  {author} {\bibinfo {author} {\bibfnamefont {G.~R.}\ \bibnamefont
  {Price}},\ }\bibfield  {title} {\enquote {\bibinfo {title} {The nature of
  selection},}\ }\href@noop {} {\bibfield  {journal} {\bibinfo  {journal}
  {Journal of Theoretical Biology}\ }\textbf {\bibinfo {volume} {175}},\
  \bibinfo {pages} {389--396} (\bibinfo {year} {1995})}\BibitemShut {NoStop}%
\bibitem [{\citenamefont {Frank}(2009)}]{frank09natural}%
  \BibitemOpen
  \bibfield  {author} {\bibinfo {author} {\bibfnamefont {S.~A.}\ \bibnamefont
  {Frank}},\ }\bibfield  {title} {\enquote {\bibinfo {title} {Natural selection
  maximizes {F}isher information},}\ }\href@noop {} {\bibfield  {journal}
  {\bibinfo  {journal} {Journal of Evolutionary Biology}\ }\textbf {\bibinfo
  {volume} {22}},\ \bibinfo {pages} {231--244} (\bibinfo {year}
  {2009})}\BibitemShut {NoStop}%
\bibitem [{\citenamefont {Feynman}(1967)}]{feynman67the-character}%
  \BibitemOpen
  \bibfield  {author} {\bibinfo {author} {\bibfnamefont {R.~P.}\ \bibnamefont
  {Feynman}},\ }\href@noop {} {\emph {\bibinfo {title} {{The Character of
  Physical Law}}}}\ (\bibinfo  {publisher} {MIT Press},\ \bibinfo {address}
  {Cambridge, MA},\ \bibinfo {year} {1967})\BibitemShut {NoStop}%
\bibitem [{\citenamefont {Weyl}(1983)}]{weyl83symmetry}%
  \BibitemOpen
  \bibfield  {author} {\bibinfo {author} {\bibfnamefont {H.}~\bibnamefont
  {Weyl}},\ }\href@noop {} {\emph {\bibinfo {title} {Symmetry}}}\ (\bibinfo
  {publisher} {Princeton University Press},\ \bibinfo {address} {Princeton,
  NJ},\ \bibinfo {year} {1983})\BibitemShut {NoStop}%
\bibitem [{\citenamefont {Fisher}(1930)}]{fisher30the-genetical}%
  \BibitemOpen
  \bibfield  {author} {\bibinfo {author} {\bibfnamefont {R.~A.}\ \bibnamefont
  {Fisher}},\ }\href@noop {} {\emph {\bibinfo {title} {The {G}enetical {T}heory
  of {N}atural {S}election}}}\ (\bibinfo  {publisher} {Clarendon},\ \bibinfo
  {address} {Oxford},\ \bibinfo {year} {1930})\BibitemShut {NoStop}%
\bibitem [{\citenamefont {Fisher}(1941)}]{fisher41average}%
  \BibitemOpen
  \bibfield  {author} {\bibinfo {author} {\bibfnamefont {R.~A.}\ \bibnamefont
  {Fisher}},\ }\bibfield  {title} {\enquote {\bibinfo {title} {Average excess
  and average effect of a gene substitution},}\ }\href@noop {} {\bibfield
  {journal} {\bibinfo  {journal} {Annals of Eugenics}\ }\textbf {\bibinfo
  {volume} {11}},\ \bibinfo {pages} {53--63} (\bibinfo {year}
  {1941})}\BibitemShut {NoStop}%
\bibitem [{\citenamefont {Crow}\ and\ \citenamefont
  {Kimura}(1970)}]{crow70an-introduction}%
  \BibitemOpen
  \bibfield  {author} {\bibinfo {author} {\bibfnamefont {J.~F.}\ \bibnamefont
  {Crow}}\ and\ \bibinfo {author} {\bibfnamefont {M.}~\bibnamefont {Kimura}},\
  }\href@noop {} {\emph {\bibinfo {title} {An {I}ntroduction to {P}opulation
  {G}enetics {T}heory}}}\ (\bibinfo  {publisher} {Burgess},\ \bibinfo {address}
  {Minneapolis, Minnesota},\ \bibinfo {year} {1970})\BibitemShut {NoStop}%
\bibitem [{\citenamefont {Ewens}(1992)}]{ewens92an-optimizing}%
  \BibitemOpen
  \bibfield  {author} {\bibinfo {author} {\bibfnamefont {W.~J.}\ \bibnamefont
  {Ewens}},\ }\bibfield  {title} {\enquote {\bibinfo {title} {An optimizing
  principle of natural selection in evolutionary population genetics},}\
  }\href@noop {} {\bibfield  {journal} {\bibinfo  {journal} {Theoretical
  Population Biology}\ }\textbf {\bibinfo {volume} {42}},\ \bibinfo {pages}
  {333--346} (\bibinfo {year} {1992})}\BibitemShut {NoStop}%
\bibitem [{\citenamefont {Frieden}\ \emph {et~al.}(2001)\citenamefont
  {Frieden}, \citenamefont {Plastino},\ and\ \citenamefont
  {Soffer}}]{frieden01population}%
  \BibitemOpen
  \bibfield  {author} {\bibinfo {author} {\bibfnamefont {B.~R.}\ \bibnamefont
  {Frieden}}, \bibinfo {author} {\bibfnamefont {A.}~\bibnamefont {Plastino}}, \
  and\ \bibinfo {author} {\bibfnamefont {B.~H.}\ \bibnamefont {Soffer}},\
  }\bibfield  {title} {\enquote {\bibinfo {title} {Population genetics from an
  information perspective},}\ }\href@noop {} {\bibfield  {journal} {\bibinfo
  {journal} {Journal of Theoretical Biology}\ }\textbf {\bibinfo {volume}
  {208}},\ \bibinfo {pages} {49--64} (\bibinfo {year} {2001})}\BibitemShut
  {NoStop}%
\bibitem [{\citenamefont {Frieden}(2004)}]{frieden04science}%
  \BibitemOpen
  \bibfield  {author} {\bibinfo {author} {\bibfnamefont {B.~R.}\ \bibnamefont
  {Frieden}},\ }\href@noop {} {\emph {\bibinfo {title} {{Science from Fisher
  Information: A Unification}}}}\ (\bibinfo  {publisher} {Cambridge University
  Press},\ \bibinfo {address} {Cambridge, UK},\ \bibinfo {year}
  {2004})\BibitemShut {NoStop}%
\bibitem [{\citenamefont {Price}(1972{\natexlab{b}})}]{price72fishers}%
  \BibitemOpen
  \bibfield  {author} {\bibinfo {author} {\bibfnamefont {G.~R.}\ \bibnamefont
  {Price}},\ }\bibfield  {title} {\enquote {\bibinfo {title} {Fisher's
  `fundamental theorem' made clear},}\ }\href@noop {} {\bibfield  {journal}
  {\bibinfo  {journal} {Annals of Human Genetics}\ }\textbf {\bibinfo {volume}
  {36}},\ \bibinfo {pages} {129--140} (\bibinfo {year}
  {1972}{\natexlab{b}})}\BibitemShut {NoStop}%
\bibitem [{\citenamefont {Ewens}(1989)}]{ewens89an-interpretation}%
  \BibitemOpen
  \bibfield  {author} {\bibinfo {author} {\bibfnamefont {W.~J.}\ \bibnamefont
  {Ewens}},\ }\bibfield  {title} {\enquote {\bibinfo {title} {An interpretation
  and proof of the fundamental theorem of natural selection},}\ }\href@noop {}
  {\bibfield  {journal} {\bibinfo  {journal} {Theoretical Population Biology}\
  }\textbf {\bibinfo {volume} {36}},\ \bibinfo {pages} {167--180} (\bibinfo
  {year} {1989})}\BibitemShut {NoStop}%
\bibitem [{\citenamefont {Frank}\ and\ \citenamefont
  {Slatkin}(1992)}]{frank92fishers}%
  \BibitemOpen
  \bibfield  {author} {\bibinfo {author} {\bibfnamefont {S.~A.}\ \bibnamefont
  {Frank}}\ and\ \bibinfo {author} {\bibfnamefont {M.}~\bibnamefont
  {Slatkin}},\ }\bibfield  {title} {\enquote {\bibinfo {title} {Fisher's
  fundamental theorem of natural selection},}\ }\href@noop {} {\bibfield
  {journal} {\bibinfo  {journal} {Trends in Ecology and Evolution}\ }\textbf
  {\bibinfo {volume} {7}},\ \bibinfo {pages} {92--95} (\bibinfo {year}
  {1992})}\BibitemShut {NoStop}%
\bibitem [{\citenamefont {Frank}(2012{\natexlab{a}})}]{frank12natural}%
  \BibitemOpen
  \bibfield  {author} {\bibinfo {author} {\bibfnamefont {S.~A.}\ \bibnamefont
  {Frank}},\ }\bibfield  {title} {\enquote {\bibinfo {title} {Natural
  selection. {III}. {S}election versus transmission and the levels of
  selection},}\ }\href@noop {} {\bibfield  {journal} {\bibinfo  {journal}
  {Journal of Evolutionary Biology}\ }\textbf {\bibinfo {volume} {25}},\
  \bibinfo {pages} {227--243} (\bibinfo {year}
  {2012}{\natexlab{a}})}\BibitemShut {NoStop}%
\bibitem [{\citenamefont {Coplien}(1998)}]{coplien98to-iterate}%
  \BibitemOpen
  \bibfield  {author} {\bibinfo {author} {\bibfnamefont {J.~O.}\ \bibnamefont
  {Coplien}},\ }\bibfield  {title} {\enquote {\bibinfo {title} {To iterate is
  human, to recurse, divine},}\ }\href@noop {} {\bibfield  {journal} {\bibinfo
  {journal} {C++ Report}\ }\textbf {\bibinfo {volume} {10}},\ \bibinfo {pages}
  {43--51} (\bibinfo {year} {1998})}\BibitemShut {NoStop}%
\bibitem [{\citenamefont {Li}(1967)}]{li67fundamental}%
  \BibitemOpen
  \bibfield  {author} {\bibinfo {author} {\bibfnamefont {C.~C.}\ \bibnamefont
  {Li}},\ }\bibfield  {title} {\enquote {\bibinfo {title} {Fundamental theorem
  of natural selection},}\ }\href@noop {} {\bibfield  {journal} {\bibinfo
  {journal} {Nature}\ }\textbf {\bibinfo {volume} {214}},\ \bibinfo {pages}
  {505--506} (\bibinfo {year} {1967})}\BibitemShut {NoStop}%
\bibitem [{\citenamefont {Fisher}(1918)}]{fisher18the-correlation}%
  \BibitemOpen
  \bibfield  {author} {\bibinfo {author} {\bibfnamefont {R.~A.}\ \bibnamefont
  {Fisher}},\ }\bibfield  {title} {\enquote {\bibinfo {title} {The correlation
  between relatives on the supposition of {{M}}endelian inheritance},}\
  }\href@noop {} {\bibfield  {journal} {\bibinfo  {journal} {Transactions of
  the Royal Society of Edinburgh}\ }\textbf {\bibinfo {volume} {52}},\ \bibinfo
  {pages} {399--433} (\bibinfo {year} {1918})}\BibitemShut {NoStop}%
\bibitem [{\citenamefont {Lynch}\ and\ \citenamefont
  {Walsh}(1998)}]{lynch98genetics}%
  \BibitemOpen
  \bibfield  {author} {\bibinfo {author} {\bibfnamefont {M.}~\bibnamefont
  {Lynch}}\ and\ \bibinfo {author} {\bibfnamefont {B.}~\bibnamefont {Walsh}},\
  }\href@noop {} {\emph {\bibinfo {title} {Genetics and {A}nalysis of
  {Q}uantitative {T}raits}}}\ (\bibinfo  {publisher} {Sinauer Associates},\
  \bibinfo {address} {Sunderland, Massachusetts},\ \bibinfo {year}
  {1998})\BibitemShut {NoStop}%
\bibitem [{\citenamefont {Hartl}(2006)}]{hartl06principles}%
  \BibitemOpen
  \bibfield  {author} {\bibinfo {author} {\bibfnamefont {D.~L.}\ \bibnamefont
  {Hartl}},\ }\href@noop {} {\emph {\bibinfo {title} {{Principles of Population
  Genetics}}}},\ \bibinfo {edition} {4th}\ ed.\ (\bibinfo  {publisher}
  {Sinauer},\ \bibinfo {address} {Sunderland, MA},\ \bibinfo {year}
  {2006})\BibitemShut {NoStop}%
\bibitem [{\citenamefont {Crow}\ and\ \citenamefont
  {Nagylaki}(1976)}]{crow76the-rate}%
  \BibitemOpen
  \bibfield  {author} {\bibinfo {author} {\bibfnamefont {J.~F.}\ \bibnamefont
  {Crow}}\ and\ \bibinfo {author} {\bibfnamefont {T.}~\bibnamefont
  {Nagylaki}},\ }\bibfield  {title} {\enquote {\bibinfo {title} {The rate of
  change of a character correlated with fitness},}\ }\href@noop {} {\bibfield
  {journal} {\bibinfo  {journal} {American Naturalist}\ }\textbf {\bibinfo
  {volume} {110}},\ \bibinfo {pages} {207--213} (\bibinfo {year}
  {1976})}\BibitemShut {NoStop}%
\bibitem [{\citenamefont {Grafen}(1984)}]{grafen84natural}%
  \BibitemOpen
  \bibfield  {author} {\bibinfo {author} {\bibfnamefont {A.}~\bibnamefont
  {Grafen}},\ }\bibfield  {title} {\enquote {\bibinfo {title} {Natural
  selection, kin selection and group selection},}\ }in\ \href@noop {} {\emph
  {\bibinfo {booktitle} {Behavioural Ecology}}},\ \bibinfo {editor} {edited by\
  \bibinfo {editor} {\bibfnamefont {J.~R.}\ \bibnamefont {Krebs}}\ and\
  \bibinfo {editor} {\bibfnamefont {N.~B.}\ \bibnamefont {Davies}}}\ (\bibinfo
  {publisher} {Blackwell Scientific Publications},\ \bibinfo {address}
  {Oxford},\ \bibinfo {year} {1984})\ pp.\ \bibinfo {pages}
  {62--84}\BibitemShut {NoStop}%
\bibitem [{\citenamefont {Russell}(1958)}]{russell58the-abc-of-relativity}%
  \BibitemOpen
  \bibfield  {author} {\bibinfo {author} {\bibfnamefont {B.}~\bibnamefont
  {Russell}},\ }\href@noop {} {\emph {\bibinfo {title} {The {ABC} of
  {R}elativity}}}\ (\bibinfo  {publisher} {New American Library},\ \bibinfo
  {address} {New York},\ \bibinfo {year} {1958})\BibitemShut {NoStop}%
\bibitem [{\citenamefont {Heisler}\ and\ \citenamefont
  {Damuth}(1987)}]{heisler87a-method}%
  \BibitemOpen
  \bibfield  {author} {\bibinfo {author} {\bibfnamefont {I.~L.}\ \bibnamefont
  {Heisler}}\ and\ \bibinfo {author} {\bibfnamefont {J.}~\bibnamefont
  {Damuth}},\ }\bibfield  {title} {\enquote {\bibinfo {title} {A method for
  analyzing selection in hierarchically structured populations},}\ }\href@noop
  {} {\bibfield  {journal} {\bibinfo  {journal} {American Naturalist}\ }\textbf
  {\bibinfo {volume} {130}},\ \bibinfo {pages} {582--602} (\bibinfo {year}
  {1987})}\BibitemShut {NoStop}%
\bibitem [{\citenamefont {Li}(1975)}]{li75path}%
  \BibitemOpen
  \bibfield  {author} {\bibinfo {author} {\bibfnamefont {C.~C.}\ \bibnamefont
  {Li}},\ }\href@noop {} {\emph {\bibinfo {title} {Path {A}nalysis}}}\
  (\bibinfo  {publisher} {Boxwood},\ \bibinfo {address} {Pacific Grove,
  California},\ \bibinfo {year} {1975})\BibitemShut {NoStop}%
\bibitem [{\citenamefont {Fox}\ and\ \citenamefont
  {Kerr}(2012)}]{fox12analyzing}%
  \BibitemOpen
  \bibfield  {author} {\bibinfo {author} {\bibfnamefont {J.~W.}\ \bibnamefont
  {Fox}}\ and\ \bibinfo {author} {\bibfnamefont {B.}~\bibnamefont {Kerr}},\
  }\bibfield  {title} {\enquote {\bibinfo {title} {Analyzing the effects of
  species gain and loss on ecosystem function using the extended {P}rice
  equation partition},}\ }\href@noop {} {\bibfield  {journal} {\bibinfo
  {journal} {Oikos}\ }\textbf {\bibinfo {volume} {121}},\ \bibinfo {pages}
  {290--298} (\bibinfo {year} {2012})}\BibitemShut {NoStop}%
\bibitem [{\citenamefont {Grafen}(1999)}]{grafen99formal}%
  \BibitemOpen
  \bibfield  {author} {\bibinfo {author} {\bibfnamefont {A.}~\bibnamefont
  {Grafen}},\ }\bibfield  {title} {\enquote {\bibinfo {title} {Formal
  {D}arwinism, the individual--as--maximizing--agent analogy and
  bet--hedging},}\ }\href@noop {} {\bibfield  {journal} {\bibinfo  {journal}
  {Proc. R. Soc. Lond. B}\ }\textbf {\bibinfo {volume} {266}},\ \bibinfo
  {pages} {799--803} (\bibinfo {year} {1999})}\BibitemShut {NoStop}%
\bibitem [{\citenamefont {Rice}(2008)}]{rice08a-stochastic}%
  \BibitemOpen
  \bibfield  {author} {\bibinfo {author} {\bibfnamefont {S.~H.}\ \bibnamefont
  {Rice}},\ }\bibfield  {title} {\enquote {\bibinfo {title} {A stochastic
  version of the {P}rice equation reveals the interplay of deterministic and
  stochastic processes in evolution},}\ }\href@noop {} {\bibfield  {journal}
  {\bibinfo  {journal} {BMC Evolutionary Biology}\ }\textbf {\bibinfo {volume}
  {8}},\ \bibinfo {pages} {262} (\bibinfo {year} {2008})}\BibitemShut {NoStop}%
\bibitem [{\citenamefont {Wolf}\ \emph {et~al.}(1998)\citenamefont {Wolf},
  \citenamefont {Brodie}, \citenamefont {Cheverud}, \citenamefont {Moore},\
  and\ \citenamefont {Wade}}]{wolf98evolutionary}%
  \BibitemOpen
  \bibfield  {author} {\bibinfo {author} {\bibfnamefont {J.~B.}\ \bibnamefont
  {Wolf}}, \bibinfo {author} {\bibfnamefont {E.~D.}\ \bibnamefont {Brodie}},
  \bibinfo {author} {\bibfnamefont {J.~M.}\ \bibnamefont {Cheverud}}, \bibinfo
  {author} {\bibfnamefont {A.~J.}\ \bibnamefont {Moore}}, \ and\ \bibinfo
  {author} {\bibfnamefont {M.~J.}\ \bibnamefont {Wade}},\ }\bibfield  {title}
  {\enquote {\bibinfo {title} {Evolutionary consequences of indirect genetic
  effects},}\ }\href@noop {} {\bibfield  {journal} {\bibinfo  {journal} {Trends
  in Ecology and Evolution}\ }\textbf {\bibinfo {volume} {13}},\ \bibinfo
  {pages} {64--69} (\bibinfo {year} {1998})}\BibitemShut {NoStop}%
\bibitem [{\citenamefont {Bijma}\ and\ \citenamefont
  {Wade}(2008)}]{bijma08the-joint}%
  \BibitemOpen
  \bibfield  {author} {\bibinfo {author} {\bibfnamefont {P.}~\bibnamefont
  {Bijma}}\ and\ \bibinfo {author} {\bibfnamefont {M.~J.}\ \bibnamefont
  {Wade}},\ }\bibfield  {title} {\enquote {\bibinfo {title} {The joint effects
  of kin, multilevel selection and indirect genetic effects on response to
  genetic selection},}\ }\href@noop {} {\bibfield  {journal} {\bibinfo
  {journal} {Journal of Evolutionary Biology}\ }\textbf {\bibinfo {volume}
  {21}},\ \bibinfo {pages} {1175--1188} (\bibinfo {year} {2008})}\BibitemShut
  {NoStop}%
\bibitem [{\citenamefont {Kahneman}(2011)}]{kahneman11thinking}%
  \BibitemOpen
  \bibfield  {author} {\bibinfo {author} {\bibfnamefont {D.}~\bibnamefont
  {Kahneman}},\ }\href@noop {} {\emph {\bibinfo {title} {{Thinking, Fast and
  Slow}}}}\ (\bibinfo  {publisher} {Farrar Straus \& Giroux},\ \bibinfo
  {address} {New York},\ \bibinfo {year} {2011})\BibitemShut {NoStop}%
\bibitem [{\citenamefont {van Veelen}(2005)}]{veelen05on-the-use-of-the-price}%
  \BibitemOpen
  \bibfield  {author} {\bibinfo {author} {\bibfnamefont {M.}~\bibnamefont {van
  Veelen}},\ }\bibfield  {title} {\enquote {\bibinfo {title} {On the use of the
  {P}rice equation},}\ }\href@noop {} {\bibfield  {journal} {\bibinfo
  {journal} {Journal of Theoretical Biology}\ }\textbf {\bibinfo {volume}
  {237}},\ \bibinfo {pages} {412--426} (\bibinfo {year} {2005})}\BibitemShut
  {NoStop}%
\bibitem [{\citenamefont {van Veelen}\ \emph {et~al.}(2010)\citenamefont {van
  Veelen}, \citenamefont {Garc{\'\i}a}, \citenamefont {Sabelis},\ and\
  \citenamefont {Egas}}]{veelen10call}%
  \BibitemOpen
  \bibfield  {author} {\bibinfo {author} {\bibfnamefont {M.}~\bibnamefont {van
  Veelen}}, \bibinfo {author} {\bibfnamefont {J.}~\bibnamefont {Garc{\'\i}a}},
  \bibinfo {author} {\bibfnamefont {M.~W.}\ \bibnamefont {Sabelis}}, \ and\
  \bibinfo {author} {\bibfnamefont {M.}~\bibnamefont {Egas}},\ }\bibfield
  {title} {\enquote {\bibinfo {title} {Call for a return to rigour in
  models},}\ }\href@noop {} {\bibfield  {journal} {\bibinfo  {journal}
  {Nature}\ }\textbf {\bibinfo {volume} {466}},\ \bibinfo {pages} {661--661}
  (\bibinfo {year} {2010})}\BibitemShut {NoStop}%
\bibitem [{\citenamefont {van Veelen}(2011)}]{van-veelen11a-rule}%
  \BibitemOpen
  \bibfield  {author} {\bibinfo {author} {\bibfnamefont {M.}~\bibnamefont {van
  Veelen}},\ }\bibfield  {title} {\enquote {\bibinfo {title} {A rule is not a
  rule if it changes from case to case (a reply to {M}arshall's comment)},}\
  }\href@noop {} {\bibfield  {journal} {\bibinfo  {journal} {Journal of
  Theoretical Biology}\ }\textbf {\bibinfo {volume} {270}},\ \bibinfo {pages}
  {189--195} (\bibinfo {year} {2011})}\BibitemShut {NoStop}%
\bibitem [{\citenamefont {Nowak}\ \emph {et~al.}(2010)\citenamefont {Nowak},
  \citenamefont {Tarnita},\ and\ \citenamefont
  {Wilson}}]{nowak10the-evolution}%
  \BibitemOpen
  \bibfield  {author} {\bibinfo {author} {\bibfnamefont {M.~A.}\ \bibnamefont
  {Nowak}}, \bibinfo {author} {\bibfnamefont {C.~E.}\ \bibnamefont {Tarnita}},
  \ and\ \bibinfo {author} {\bibfnamefont {E.~O.}\ \bibnamefont {Wilson}},\
  }\bibfield  {title} {\enquote {\bibinfo {title} {The evolution of
  eusociality},}\ }\href@noop {} {\bibfield  {journal} {\bibinfo  {journal}
  {Nature}\ }\textbf {\bibinfo {volume} {466}},\ \bibinfo {pages} {1057--1062}
  (\bibinfo {year} {2010})}\BibitemShut {NoStop}%
\bibitem [{\citenamefont {Lewontin}(1974)}]{lewontin74the-genetic}%
  \BibitemOpen
  \bibfield  {author} {\bibinfo {author} {\bibfnamefont {R.~C.}\ \bibnamefont
  {Lewontin}},\ }\href@noop {} {\emph {\bibinfo {title} {The {G}enetic {B}asis
  of {E}volutionary {C}hange}}}\ (\bibinfo  {publisher} {Columbia University
  Press},\ \bibinfo {address} {New York},\ \bibinfo {year} {1974})\BibitemShut
  {NoStop}%
\bibitem [{\citenamefont {Russell}(1922)}]{russell22introduction}%
  \BibitemOpen
  \bibfield  {author} {\bibinfo {author} {\bibfnamefont {B.}~\bibnamefont
  {Russell}},\ }\href@noop {} {\emph {\bibinfo {title} {{Introduction to
  Tractatus Logico-Philosophicus by Ludwig Wittgenstein}}}}\ (\bibinfo
  {publisher} {Harcourt, Brace and Company},\ \bibinfo {address} {New York},\
  \bibinfo {year} {1922})\BibitemShut {NoStop}%
\bibitem [{\citenamefont {Hardy}(1967)}]{hardy67a-mathematicians}%
  \BibitemOpen
  \bibfield  {author} {\bibinfo {author} {\bibfnamefont {G.~H.}\ \bibnamefont
  {Hardy}},\ }\href@noop {} {\emph {\bibinfo {title} {{A Mathematician's
  Apology}}}}\ (\bibinfo  {publisher} {Cambridge Univerity Press},\ \bibinfo
  {address} {Cambridge, UK},\ \bibinfo {year} {1967})\BibitemShut {NoStop}%
\bibitem [{\citenamefont {Frank}(2012{\natexlab{b}})}]{frank12wrights}%
  \BibitemOpen
  \bibfield  {author} {\bibinfo {author} {\bibfnamefont {S.~A.}\ \bibnamefont
  {Frank}},\ }\bibfield  {title} {\enquote {\bibinfo {title} {Wright's adaptive
  landscape versus {F}isher's fundamental theorem},}\ }in\ \href
  {http://arxiv.org/abs/1102.3709v1} {\emph {\bibinfo {booktitle} {{The
  Adaptive Landscape in Evolutionary Biology}}}},\ \bibinfo {editor} {edited
  by\ \bibinfo {editor} {\bibfnamefont {E.}~\bibnamefont {Svensson}}\ and\
  \bibinfo {editor} {\bibfnamefont {R.}~\bibnamefont {Calsbeek}}}\ (\bibinfo
  {publisher} {Oxford University Press},\ \bibinfo {address} {New York},\
  \bibinfo {year} {2012})\ p.\ \bibinfo {pages} {(in press)}\BibitemShut
  {NoStop}%
\bibitem [{\citenamefont {Anderson}(1972)}]{anderson72more}%
  \BibitemOpen
  \bibfield  {author} {\bibinfo {author} {\bibfnamefont {P.}~\bibnamefont
  {Anderson}},\ }\bibfield  {title} {\enquote {\bibinfo {title} {More is
  different},}\ }\href@noop {} {\bibfield  {journal} {\bibinfo  {journal}
  {Science}\ }\textbf {\bibinfo {volume} {177}},\ \bibinfo {pages} {393--396}
  (\bibinfo {year} {1972})}\BibitemShut {NoStop}%
\bibitem [{\citenamefont {Frank}\ and\ \citenamefont
  {Smith}(2010)}]{frank10measurement}%
  \BibitemOpen
  \bibfield  {author} {\bibinfo {author} {\bibfnamefont {S.~A.}\ \bibnamefont
  {Frank}}\ and\ \bibinfo {author} {\bibfnamefont {E.}~\bibnamefont {Smith}},\
  }\bibfield  {title} {\enquote {\bibinfo {title} {Measurement invariance,
  entropy, and probability},}\ }\href@noop {} {\bibfield  {journal} {\bibinfo
  {journal} {Entropy}\ }\textbf {\bibinfo {volume} {12}},\ \bibinfo {pages}
  {289--303} (\bibinfo {year} {2010})}\BibitemShut {NoStop}%
\bibitem [{\citenamefont {Frank}\ and\ \citenamefont
  {Smith}(2011)}]{frank11a-simple}%
  \BibitemOpen
  \bibfield  {author} {\bibinfo {author} {\bibfnamefont {S.~A.}\ \bibnamefont
  {Frank}}\ and\ \bibinfo {author} {\bibfnamefont {E.}~\bibnamefont {Smith}},\
  }\bibfield  {title} {\enquote {\bibinfo {title} {A simple derivation and
  classification of common probability distributions based on information
  symmetry and measurement scale.}}\ }\href@noop {} {\bibfield  {journal}
  {\bibinfo  {journal} {Journal of Evolutionary Biology}\ }\textbf {\bibinfo
  {volume} {24}},\ \bibinfo {pages} {469--484} (\bibinfo {year}
  {2011})}\BibitemShut {NoStop}%
\bibitem [{\citenamefont {Frank}(2011)}]{frank11measurement}%
  \BibitemOpen
  \bibfield  {author} {\bibinfo {author} {\bibfnamefont {S.~A.}\ \bibnamefont
  {Frank}},\ }\bibfield  {title} {\enquote {\bibinfo {title} {Measurement scale
  in maximum entropy models of species abundance},}\ }\href@noop {} {\bibfield
  {journal} {\bibinfo  {journal} {Journal of Evolutionary Biology}\ }\textbf
  {\bibinfo {volume} {24}},\ \bibinfo {pages} {485--496} (\bibinfo {year}
  {2011})}\BibitemShut {NoStop}%
\bibitem [{\citenamefont {Houle}\ \emph {et~al.}(2011)\citenamefont {Houle},
  \citenamefont {P{\'e}labon}, \citenamefont {Wagner},\ and\ \citenamefont
  {Hansen}}]{houle11measurement}%
  \BibitemOpen
  \bibfield  {author} {\bibinfo {author} {\bibfnamefont {D.}~\bibnamefont
  {Houle}}, \bibinfo {author} {\bibfnamefont {C.}~\bibnamefont {P{\'e}labon}},
  \bibinfo {author} {\bibfnamefont {G.~P.}\ \bibnamefont {Wagner}}, \ and\
  \bibinfo {author} {\bibfnamefont {T.~F.}\ \bibnamefont {Hansen}},\ }\bibfield
   {title} {\enquote {\bibinfo {title} {Measurement and meaning in biology},}\
  }\href@noop {} {\bibfield  {journal} {\bibinfo  {journal} {Quarterly Review
  of Biology}\ }\textbf {\bibinfo {volume} {86}},\ \bibinfo {pages} {3--34}
  (\bibinfo {year} {2011})}\BibitemShut {NoStop}%
\end{thebibliography}%

\end{document}